\documentclass[journal,10pt,compsoc]{IEEEtran}

\usepackage{multirow}
\usepackage{graphicx,comment,dblfloatfix} 
\usepackage{tabularx}
\usepackage{url}
\usepackage{color}
\usepackage{epstopdf}
\usepackage{autonum}
\usepackage{enumitem}
\usepackage{dsfont,amssymb}

\usepackage[algoruled,medskip,dontprintsemicolon,linesnumbered,Algorithm]{algorithm}
\usepackage[noend]{algpseudocode}

\newcommand{\Fig}[1]{Fig.~\ref{fig:#1}}
\newcommand{\Sec}[1]{Sec.~\ref{sec:#1}}
\newcommand{\Tab}[1]{Tab.~\ref{tab:#1}}
\newcommand{\Eq}[1]{(\ref{eq:#1})}
\newcommand{\Alg}[1]{Alg.~\ref{alg:#1}}
\newcommand{\Line}[1]{Line~\ref{line:#1}}

\newcommand{\Prop}[1]{Proposition~\ref{prop:#1}}

\newcommand{\Cor}[1]{Corollary~\ref{cor:#1}}

\newtheorem{proposition}{Proposition}
\newtheorem{example}{Example}
\newtheorem{theorem}{Theorem}
\newtheorem{corollary}{Corollary}

\newcommand{\Qc}{\mathcal{Q}}
\newcommand{\Kc}{\mathcal{K}}

\newcommand{\Wc}{\mathcal{W}}

\newcommand{\PP}{\mathds{P}} %

\begin{document}

\title{
Edge-powered Assisted Driving\\For Connected Cars
}

\author{Francesco~Malandrino,~\IEEEmembership{Senior~Member,~IEEE,}
Carla~Fabiana~Chiasserini,~\IEEEmembership{Fellow,~IEEE,}
and~Gian~Michele~Dell'Aera
\thanks{F.~Malandrino and C.~F.~Chiasserini are with CNR-IEIIT and CNIT, Italy. C.~F.~Chiasserini is with Politecnico di Torino and CARS Lab, Italy. G.~M.~Dell'Aera is with TIM, Italy.}
\thanks{This work was partially supported by TIM through the contract ``Performance Analysis of Edge Solutions 
for Automotive Applications'', and partially by the EU Commission under the RAINBOW project (Grant Agreement no. 871403).}
} %
\maketitle

\begin{abstract}
Assisted driving for connected cars is one of the main applications that 5G-and-beyond networks shall support. In this work, we propose an assisted driving system leveraging the synergy between connected vehicles and the edge of the network infrastructure, in order to envision global traffic policies that can effectively drive local decisions. Local decisions concern individual vehicles, e.g., which vehicle should perform a lane-change manoeuvre and when; global decisions, instead, involve whole traffic flows. Such decisions are made at different time scales by different entities, which are integrated within an edge-based architecture and can share information. In particular, we leverage a queuing-based model and formulate an optimization problem to make global decisions on traffic flows. To cope with the problem complexity, we then develop an iterative, linear-time complexity algorithm called Bottleneck Hunting (BH). We show the performance of our solution using a realistic simulation framework, integrating a Python engine with ns-3 and SUMO, and considering two relevant services, namely, lane change assistance and navigation, in a real-world scenario. Results demonstrate that our solution leads to a reduction of the vehicles' travel times by 66\% in the case of lane change assistance and by 20\% for navigation, compared to traditional, local-coordination approaches.
\end{abstract}


\section{Introduction}
\label{sec:intro}

5G-and-beyond networks are expected to support critical mobile
services requiring ultra-low latency and ultra-high reliability. Among
these, automotive services are pivotal to the increase of road safety and
traffic efficiency, promising a substantial reduction in
terms of car accidents and road congestion, as well as a significant improvement for 
the environment and in
both people's life quality~\cite{who-roads} and health.

In particular, both the number of accidents and traffic congestion can
be effectively decreased by enabling vehicles to {\em coordinate} their
movements with each other. Techniques for 
vehicle coordination and their impact have been investigated in various scenarios and
targeting different use cases, ranging from 
lane change and lane merge in highways \cite{zhou2016impact} to 
collision avoidance at intersections~\cite{lehoczky1972traffic} and traffic light
management~\cite{dunne1967traffic}. More recently, the advent of connected and
automated vehicles, equipped with sensing, computing, and
communication capabilities, has
spurred a dramatic revival of interest in assisted driving. 
Examples confirming this trend include the  activity  within
research projects like AutoNet2030~\cite{autonet30-commag}, 5GCAR~\cite{5gcar}, and 
5G-TRANSFORMER~\cite{5gt}, as well as the one by 
standardization organizations focusing  on cooperative intelligent transportation  
services~\cite{etsi-2019}.

In spite of the above research efforts aimed at developing effective assisted
driving systems, several challenges still have to
be solved. In particular, all existing solutions (including the 
standardized ones) leverage {\em local}
decisions (i.e., made by each {\em ego} vehicle), based on the information
collected from neighboring vehicles. Although being locally made, such
decisions have an impact on the vehicle traffic flow as a whole, hence, on
the behavior of vehicles that may be even very  far away from the ego
one.

\begin{figure}[t]
\centering
\includegraphics[width=0.9\columnwidth]{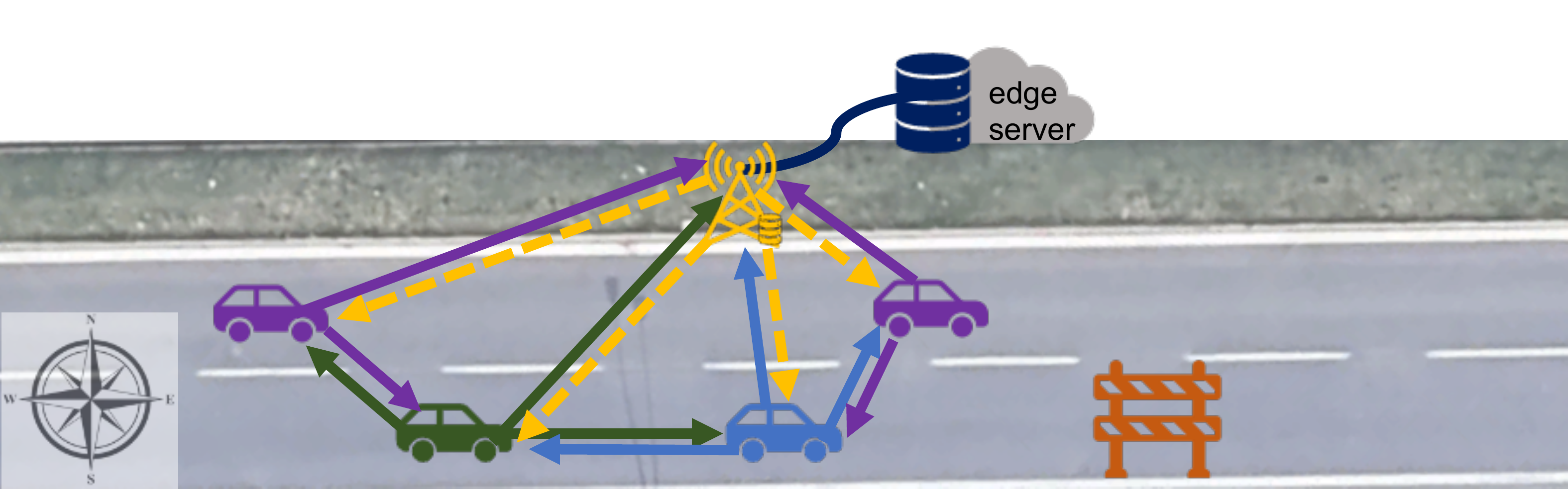}
\caption{Lane change
in a two-lane, one-way road:
the south lane is blocked  and
the two vehicles traveling on it, i.e., the blue and green ones, must move to the north lane. 
As foreseen by~\cite{etsi-2019}, vehicles announce their situation (position, speed, heading...) via CAM messages (solid lines), received by both RSUs and nearby vehicles. 
Based on such information, an edge server computes the optimal
policy  and sends it to a road-side actualizer (highlighted in yellow). 
The latter translates it into individual instructions, transmitted to
each vehicle  (dotted arrows). 
\label{fig:archi}
} %
\end{figure}

Predicting and controlling the consequences of
local decisions is a daunting task. To face such a problem, we envision a system architecture,
first introduced in our conference paper~\cite{noi-globecom},
that exploits the
synergy between the vehicular network and the edge of the network
infrastructure.  Such architecture comprises two logical layers, 
one concerned with vehicular flows and in charge of defining suitable strategies to 
optimize them,  the other addressing the movement of the individual vehicles
on a shorter time span and generating 
instructions for specific manoeuvres.
More in detail, we envision a system including:
\begin{itemize}
\item an edge server (e.g., located within  the city metro node),
     which, leveraging existing queuing-based models,  
     computes
     {\em traffic flow policies},
     for each vehicle flow, aiming at reducing the flow
    travel time;
    \item a set of servers in the radio access
      network, e.g.,  residing at cellular base stations or
      road-side units (RSUs), which translate such policies into {\em
        individual instructions}
      for the single vehicles. These servers, also referred to as
 {\em actualizers} in the rest of the paper, are in charge of delivering the
      individual instructions to the vehicles 
      so as to match the aforementioned flow policies on the longer run;
\item connected and automated vehicles, who receive the above
  instructions and  perform the suggested manoeuvre by
coordinating with their neighbors through vehicle-to-vehicle (V2V) communications.
\end{itemize}
An example of the above architecture referring to  the lane change use
case is presented in 
\Fig{archi}. Here, a policy may instruct 30\% of the cars on the
south lane to  move left early on, 
i.e., as early as where the green car is,
and the others to change lane later on, i.e., where the blue car is. 
To enact this 
policy, the road-side actualizer  instructs  the  approaching cars accordingly, 
and the vehicles receiving the lane change command interact with their
neighbors, following, e.g., the protocol  in \cite{autonet30-commag}, 
and perform the
manoeuvre.
Notice how the actualizer does not change the policies decided by the edge server, but it enforces them by translating them into local, vehicle-specific instructions. At the same time, policies do not compel the actualizer to choose any particular action for any particular vehicle; 
indeed, the edge server does not have any visibility over individual vehicles 
-- which makes computing policies easier and faster.

The communication between the network infrastructure (and
the servers therein) and the vehicles, as well
as V2V communications, occur using  standard
protocols and messages (e.g.,~\cite{etsi-2019}). Nevertheless, since the
individual instructions conveyed by the actualizer to the vehicles
realize the policies computed by the edge server, the system can
attain local objectives (e.g., avoiding
car accidents) as well as global ones (e.g., reducing traffic
congestion).

The effectiveness of the above architecture and information flow, however,
depend upon the quality of the traffic flow policies.
Indeed, the task of
computing optimal policies at the edge server is still very challenging,
due to the following reasons:
\begin{itemize} 
\item[{\em (i)}] policies have to account for the different
  destinations of the vehicles currently traveling over the same 
stretch of road, and for how the behavior of such vehicles can affect each other's travel time;  
\item[{\em (ii)}] to attain fair policies, it is essential to consider
 the vehicles travel time 
  distribution, rather than relaying  on the average  travel time only;  
\item[{\em (iii)}] in addition to being as effective as possible, policies must be
  computed efficiently so as to provide real-time driving assistance. 
\end{itemize} 

In this paper,
we tackle the above issues by providing the following main contributions:

\noindent 
{\em System architecture:} we design a system architecture,
based on the edge computing paradigm~\cite{etsi-mec-wp,rev1,rev1b,falko-vulnerable,falko-ica},
      integrating  connected cars and the edge of the network
      architecture, with the latter hosting a traffic controller in
      charge of  providing traffic  flow policies (\Sec{archi});

\noindent
 {\em Analytical modeling and optimization:}
leveraging existing works on queue-based models for vehicular traffic (see \Sec{relwork} next), 
we present in \Sec{model} a model that can
represent  arbitrary road topologies as
      well as any  path followed by the vehicles, and which allows for the computation
      of the vehicles' travel times
{\em distribution} (as opposed to a simple average), as detailed in \Sec{times}. Through this model, 
we derive the distribution of the vehicles' travel times, and
      formulate the problem of optimizing such distribution (\Sec{problem});
    
    \noindent
    {\em Algorithm:} we present and analyze an optimal algorithm,
inspired by gradient-descent and
      named {\em Bottleneck Hunting} (BH),  that provides optimal traffic
      flow policies in linear time (\Sec{algo});  

\noindent
 {\em Realistic simulation framework and performance evaluation:}
  we leverage a simulation framework including a Python engine we developed, the ns-3 network
simulator, and the SUMO mobility simulator. Through such simulator, we capture 
the main aspects of real-world assisted driving systems (\Sec{validation})
and assess the 
performance of the BH algorithm in  real-world use cases, showing
the gain we can obtain with respect to traditional, distributed strategies (\Sec{results}). 

\noindent
Finally, we conclude the paper in \Sec{conclusion}.

\section{Related work}
\label{sec:relwork}

Queuing theory has often been used for mesoscopic models of vehicular traffic, 
on both individual road stretches and more complex road layouts. 
In such models, customers in queues represent the vehicles traversing the road 
topology~\cite{queue-survey,rev3-1,rev3-2,Gomes-journal}, while, in the case of straight road stretches, 
queue service times model  the vehicle travel times.

Owing to the power and flexibility of the queuing abstraction, queues
have also been consistently and effectively used to represent
congested traffic conditions~\cite{queue-survey}
-- indeed, queuing theory can well address the causes and effects of congestion.
The pioneering work in 
\cite{alfa1995modelling}~argues that the arrival of vehicles  at intersections,
either individually or in batches, can be represented as a Poisson process.
The
study in \cite{heidemann1996queueing} proposes to model urban
topologies as a set of elementary (namely, $M/M/1$ and $M/D/1$) queues, and
the later work~\cite{van2006empirical} leverages empirical validation to argue for a more general, 
$M/G/1$ model in non-congested scenarios. In both cases, individual lanes are each associated 
with a queue.
The later work~\cite{rev3-2} finds that, under free-flow conditions, the probability that any 
road segment attains its full capacity is zero, therefore, it is appropriate to model such 
segments through queues with an infinite number of servers, namely, of type~$M/G/\infty$.

Vehicular mobility is modeled through $M/M/1$ queues also in several recent works about edge and 
cloud computing in vehicular networks, including~\cite{zhou2018begin,ning2019vehicular,zhao2018joint}.
Similarly, the work~\cite{rev3-1} leverages Poisson arrivals to characterize the connectivity 
of VANETs in highway-like scenarios.

Other works, e.g.,~\cite{mukhopadhyay2017approximate},
focus on specific aspects of vehicular traffic, such as modeling the
delay of indisciplined traffic
(where motorbikes and cars do not
respect lanes) at regulated intersections
traffic~\cite{mukhopadhyay2017approximate}, or on estimating the error
incurred by probe vehicles~\cite{comert2011analytical}. In both cases,
$M/M/1$ models are found to accurately represent real-world
conditions. Similarly, \cite{mukhopadhyay2017approximate}~shows that indisciplined
traffic can be modeled via $M/M/1$ queues, and
\cite{pisano2010mitigating}~uses them to study congestion in vehicular networks.
Consistently, the study in~\cite{carlo-secon12}, based on
realistic simulations, finds $M/M/1$ to well represent travel time on
individual road segments.
Note that part of the popularity of Markovian service times is due to their
ability to capture the effect of heterogeneous traffic (e.g., cars, trucks,...)~\cite{mukhopadhyay2017approximate}, proceeding at different speeds. 
Batch arrival and departure processes are the modeling tool of choice
when traffic signals are involved,
e.g.,~\cite{timmerman2019platoon,harahap2019modeling}. In
all cases, arrivals are assumed to be Markovian in nature, while
service times are either Markovian or deterministic.

Situations where vehicles from multiple flows have to {\em merge} into
one lane has been identified as a major cause of
congestion. Accordingly, several research efforts focus on
modeling the merging behavior of
vehicles~\cite{kanagaraj2010modeling},
as well as on devising optimized merging
strategies~\cite{wang2009proactive}.
Among merging situations, several
works~\cite{milanes2010automated} focus on access
ramps to highways, which are especially critical because of the
different speeds of incoming vehicles. Other works,
e.g.,~\cite{abbas2013radio}, focus on communication aspects,
especially the channel conditions experienced by merging vehicles in
urban scenarios.
Recent efforts such as~\cite{zhou2016impact} deal with merging strategies for autonomous vehicles, and the benefits coming from cooperation between them.

Traffic accidents, along with their causes and consequences, have long
been studied in the literature. As an example,
\cite{hojati2013hazard}~aims at characterizing the duration of
accidents and the factors that could mitigate their consequences,
e.g., better infrastructure maintenance. Closer to one of
  the scenarios we consider, \cite{baykal2009modeling}~considers the
case of a multi-lane road, where one of the lanes is blocked by an
accident. The authors of~\cite{baykal2009modeling}, however, model the
whole road as a single $M/M/c$ queue,
that is, with multiple servers ($c>1$) drawing from the same queue. 
Conversely, beside having a different scope, we
model a multiple-lane road stretch through
 multiple one-server queues.

Focusing on the specific aspect of lane-change behavior, \cite{duan2013emergency}~combines queuing theory, 
to describe the mobility of vehicles, and game theory, to model the self-interested decisions they make. 
With a different goal but using the same tools, \cite{bi2015detecting}~leverages queuing theory 
to {\em detect} the users most likely to attempt a lane change, before they do so.

The general problem of harmonizing and optimizing multiple flows of
vehicles through a road network has received
considerable attention from the research
community. %
In particular, \cite{lee2019dynamic}~envisions switching between flow-
and phase-based control of traffic lights based on real-time traffic
conditions. \cite{zhao2018joint}, instead,  accounts for the available radio 
coverage while making flow-routing decisions, to ensure that each vehicle gets the bandwidth it needs. Taking a dual approach, \cite{azizian2016optimized}~aims at identifying clusters of vehicles traveling together and using them as ``vehicular clouds'' to perform computation tasks.
Early works focus on individual
intersections and study how to model their behavior through
queuing-based models. For
instance, such models are leveraged in~\cite{dunne1967traffic} to
program an intelligent traffic light reacting to traffic conditions. 

Among recent solutions leveraging edge computing in intelligent transportation systems (ITS) scenarios, 
many focus on providing safety-critical services like collision avoidance via edge 
servers~\cite{falko-vulnerable,falko-ica}, with special emphasis on vulnerable users 
such as pedestrians and bicycle riders~\cite{falko-vulnerable}. 
A more general approach is adopted in~\cite{rev1,rev1b}, targeting the problem of placing 
any ITS service leveraging both edge and cloud resources.

Finally, a preliminary version of this work has been presented in the conference 
paper~\cite{noi-globecom}. With respect to~\cite{noi-globecom}, this paper greatly expands 
the scope of the model (which now includes batch arrivals), the characterization of the problem 
(even when no closed-form solution exists), the analysis of 
the algorithm (which we prove to have linear complexity), 
as well as the results (which we expand from one to three reference scenarios).

{\bf Novelty.} 
In this paper, we leverage existing works on queue models for vehicular networks and 
existing edge-based architectures in order to provide high-quality assistance to connected 
vehicles. Specifically, unlike existing works, we drive local decisions 
through vehicle flow policies, which account 
for the global effects of local behaviors. 
Thus, while other edge-based approaches foresee only one decision-making entity, 
in our solution global and local decisions are made by different entities, 
integrated into an edge-based architecture and able to operate at different time scales, 
while sharing information and policies. 
Furthermore, our decision-making scheme accounts for the whole distribution 
of flow-wise trip-times, as opposed to simple averages thereof.  
We prove that such a scheme is scalable 
under a wide variety of modeling assumptions and scenarios. 
Finally, our algorithm outperforms generic optimization algorithms  
by exploiting problem-specific information to further speed-up the decision process.

\section{System architecture}
\label{sec:archi}

We now introduce
the architecture we envision,
highlighting how it is
consistent with existing standards on
connected vehicle communications.  
According to the ETSI 102.941 (2019) standard, connected vehicles
periodically (e.g., every 100 ms) broadcast cooperative awareness messages (CAMs),
including their location, speed, and heading. Such information is
then used by other vehicles and/or the network infrastructure for
several safety and convenience 
services, e.g., collision avoidance~\cite{noi-vtm}. Upon detecting a
situation warranting action (e.g., two vehicles on a collision course), decentralized environmental
  notification messages (DENMs)  are sent to the affected
vehicles, so that their actualizer is triggered and/or their driver
is warned. Both CAMs and DENMs can carry additional information~\cite{etsi-2019} for the support of  assisted driving services such
as lane change/merge \cite{etsi-2019} and navigation services (see ETSI  102.638).

\begin{figure}
\centering
\includegraphics[width=0.75\columnwidth]{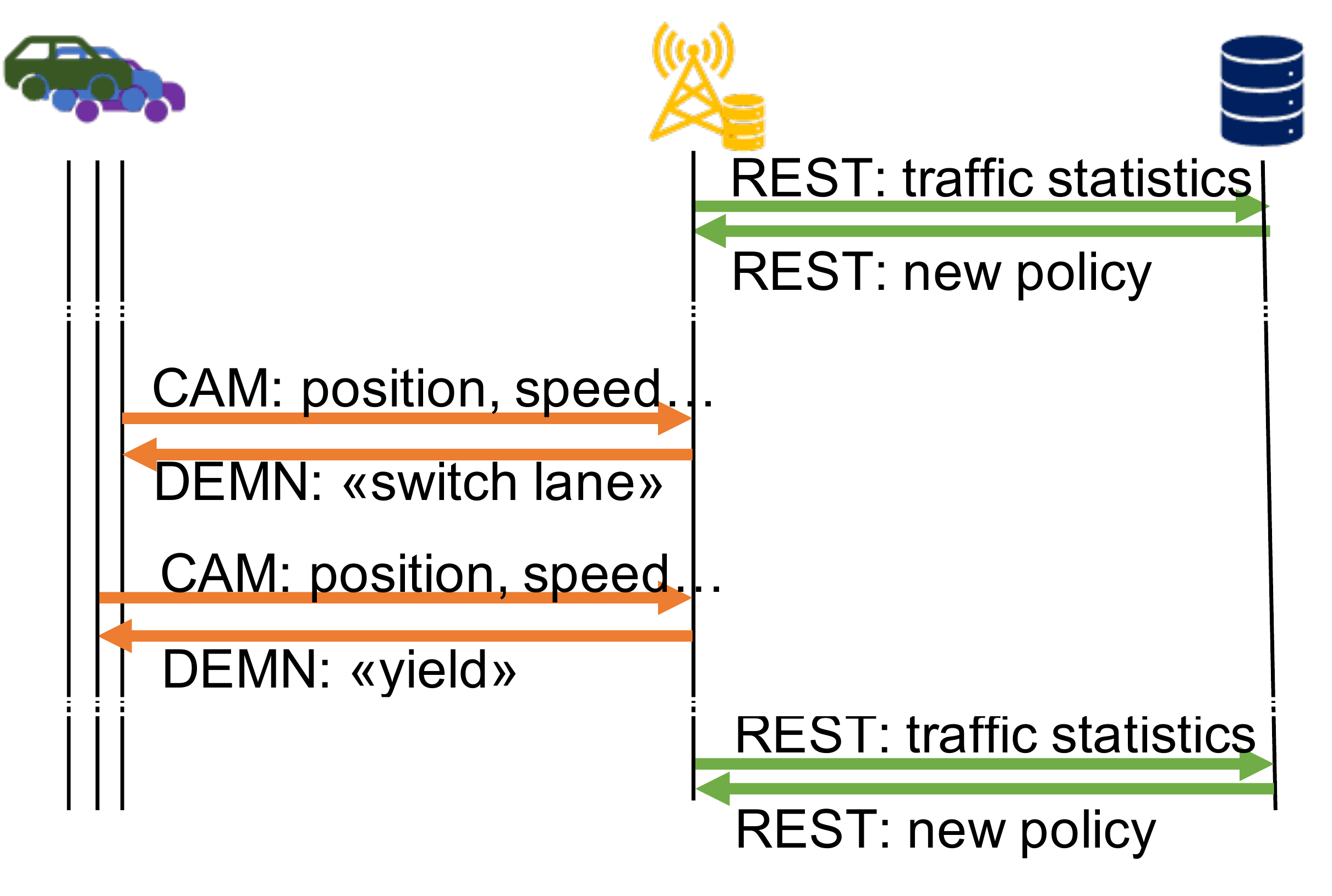}
\caption{
Lane-changing scenario: messages exchanged between vehicles, RSUs (acting as actualizers), 
and edge servers. Orange lines correspond to ETSI 102.941-standardized messages, 
green lines to HTTP-based REST exchanges. 
\label{fig:messages}
} %
\end{figure}

None of the above is envisioned to change under our proposed
architecture; in particular, all services still leverage the transmission of 
CAMs and DENMs.  For concreteness, we focus on  the lane change assistance
and navigation services.  
In the first case, the edge server exploits
the information in the CAMs received by the radio access nodes (and
possibly notifications of lane blocks carried by DENMs) to formulate  an
optimal policy. Such a policy is then sent to the actualizers, which,
being closer to the mobile users, can account for the most recent CAMs
and exploit them to  translate the policy  into individual
instructions for the single vehicles (see \Fig{archi}). These instructions are notified
to vehicles through DENMs sent by the radio access nodes. Upon
receiving a DENM, a vehicle starts interacting via V2V communications with its neighbors
following, e.g., the protocol defined in \cite{autonet30-commag} and
using CAMs and DENMs as foreseen by \cite{etsi-2019}.  

The second service (navigation) operates on a longer time and larger geographical  scale: the edge server exploits the CAMs
and, in particular, the route destination field therein, and computes
the optimal vehicles' route. This is then notified by the radio
access nodes using DENMs. 

\Fig{messages} exemplifies the messages between vehicles, RSUs (acting as actualizers), 
and the edge server in the lane-changing scenario. The vehicles and the RSU exchange 
messages standardized by ETSI 102.941, e.g., CAMs and DEMNs, represented by orange lines in 
\Fig{messages}. CAMs convey information on (respectively) the vehicle's status, 
including its position, speed, heading, acceleration, and intentions (e.g., turn signals); DEMN messages, instead, convey indications on what vehicles should do, e.g., switch lane.

RSUs and edge server, instead, communicate through HTTP-based REST exchanges, corresponding to green lines in the figure: RSUs periodically send reports on the traffic they observe, and the edge server leverages such information to periodically issue new general vehicle flow policies. RSUs account for the currently-active policies when sending out DEMN messages, e.g., to decide how many vehicles to instruct to yield.

Notice that RSUs are not {\em required} to communicate directly; however, 
if such an option is available, it can be exploited. Specifically, one of the 
RSU servers can also function as a policy server, and therefore communicate directly with other 
RSUs to distribute policies. Indeed, {\em policy maker} and {\em actualizer} are logical roles, 
which can be assigned to any physical node, provided it has the required capabilities.

For all services, latency is a foremost concern; indeed, carefully-crafted policies and 
individualized suggestions 
are no use if they come too late. Edge-based solutions such as ours have longer latency than purely 
V2V ones. 
However, previous works on collision avoidance~\cite{falko-vulnerable,falko-ica} have shown 
that such additional delay is negligible compared to the overall reaction times, 
and more than compensated by the additional information that can be leveraged. 
It is worth noticing that collision-avoidance services have extremely tight latency requirements, 
thus, the findings of~\cite{falko-vulnerable,falko-ica} also apply to, e.g., lane-changing assistance.  

Placing the server at the network edge yields the optimal trade-off between the availability of global 
information and the need for low latency. As discussed earlier, edge servers have remarkably 
short latencies, compatible with critical safety services. At the same time, they oversee fairly large 
areas, e.g., a neighborhood or a small town; it follows that the information they collect is sufficiently 
general to allow high-quality, global decisions. If the service (e.g., navigation) and the scenario 
(e.g., large urban areas) warrant it, multiple edge servers can coordinate and synchronize via the Internet cloud.

Throughout this paper, for simplicity, we refer to a {\em single} edge server. 
However, it is important to highlight that, for very complex and/or large scenarios, {\em multiple} 
edge servers can coexist, each responsible for a subset of the topology. 
This allows the edge servers to make decisions more efficiently, while retaining their ability to 
account for their far-reaching effects. Importantly, different services require edge servers to 
supervise areas of different sizes: for a lane change service, the area supervised by a 
single edge server can be as small as a single stretch of road; for navigation services, 
it can be as large as a neighborhood or small town. In the latter case, however, decisions can be made less frequently, hence, the load on the edge server remains manageable.

It is also important to highlight how, as also exemplified in \Fig{messages}, actualizers and policy 
servers operate at different {\em time scales}. Indeed, actualizers can immediately react to messages 
coming from vehicles according to the current policy, without the need for the policy server to intervene. 
Through such a logical separation between policy servers and actualizers, our solution is able to 
leverage global knowledge when making local decisions.

\section{Model design}
\label{sec:model}

In this section,  
we describe how our system model represents the road
layout, the vehicle flows (\Sec{sub-queues}), the routes taken by
vehicles in the same flow, and their travel times
(\Sec{sub-paths}). With the aim  to devise efficient policies at the edge server, 
we represent traffic flows 
by queuing models, so as to capture the 
flow dynamics depending on aggregate quantities such as the road capacity.
Importantly, 
\begin{itemize}
    \item based on the existing works and validation studies~\cite{heidemann1996queueing,van2006empirical,zhou2018begin,ning2019vehicular,zhao2018joint,rev3-1}, we consider Markovian arrivals, of either individual vehicles or  batches thereof;
    \item we do not restrict ourselves to a specific service time distribution
or number of servers,
supporting instead a very wide range of queuing models;
    \item given the scope of our work, we focus on  uncongested
      scenarios where the number of vehicles traveling on a road stretch
      is lower than its capacity, hence road stretches can be modeled with queues with
      infinite length. %
\end{itemize}

The decisions to be made correspond, intuitively, to the policies
formulated 
by the edge server in  \Fig{archi}. Specifically, they
consist of  the suggested travel speed  and the probabilities of
taking a given lane/stretch of road. In the following,
we refer to a single lane on a stretch of road as {\em road segment}.

\begin{figure}
\centering
\includegraphics[width=0.75\columnwidth]{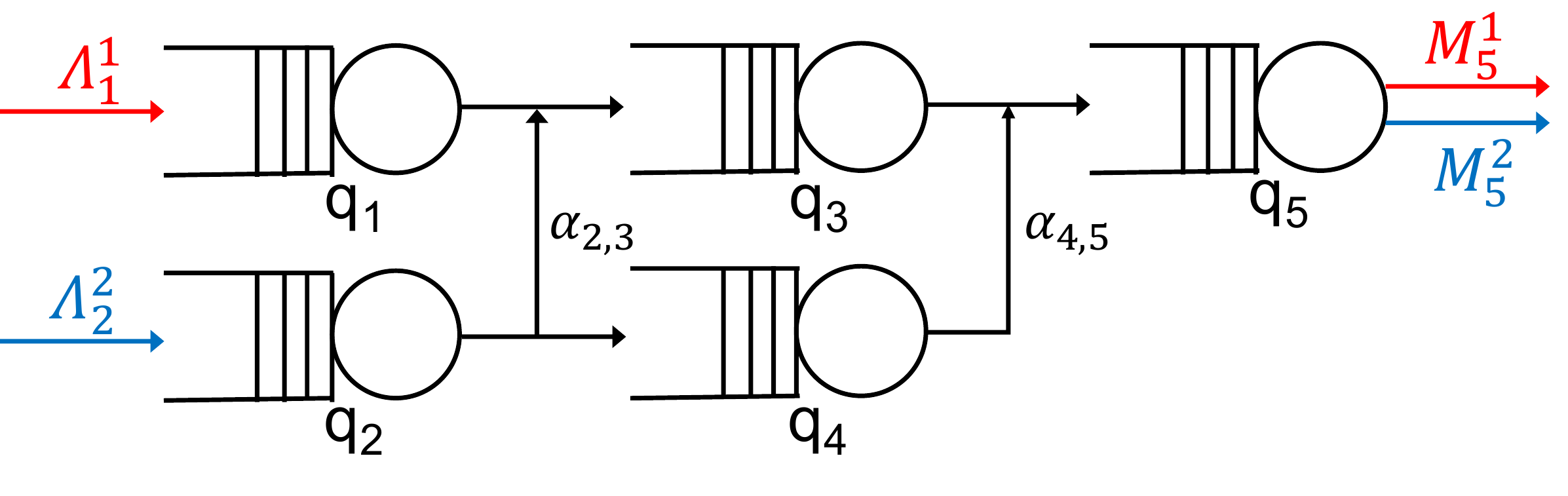}
\caption{Toy scenario: flows of vehicular traffic  
go through a set of interconnected queues, 
each representing a single-lane stretch of road.
\label{fig:scenario}
} %
\end{figure}

\subsection{Road topologies as queue networks}
\label{sec:sub-queues}

Similar to existing works~\cite{rev1,rev1b,rev3-1,rev3-2}, we
 model the road topology under the control of the edge server as a set of interconnected 
  queues~$q_i\in\Qc$, as exemplified in \Fig{scenario}. 
Similarly to existing works (including experimental validation~\cite{van2006empirical,rev3-2}), 
each queue represents a road segment, characterized by, e.g., a certain speed
limit and capacity.

Service rates,  $\mu_i$, are bounded by a maximum
value~$\mu^{\max}_i$, determined by the length of the road
segment modeled by queue $q_i$, its speed limit, and the
inter-vehicle safety distance: 
\begin{equation}
\label{eq:mu-max}
\mu_i\leq\mu^{\max}_i\quad\forall q_i\in\Qc \,.
\end{equation}

Several traffic {\em flows}, $k\in\Kc$, may
travel across the road topology or part of it, each flow being defined as a 
set of vehicles having the same source and
destination within the road topology. 
Specifically, given a flow $k$ entering and leaving the road topology
at queue $q_i$ and $q_o$, respectively, the parameter~$\Lambda^k_i$
denotes the vehicles 
arrival rate, 
while parameter~$M^k_o$ denotes the rate at which they
exit the  topology. 
For every flow~$k$, there is 
exactly one queue~$i$ for which~$\Lambda^k_i>0$ and one queue~$o$ for
which~$M^k_o>0$ (see \Fig{scenario});
furthermore, these two quantities are the same, as all vehicles of the 
flow begin their route at queue~$i$ and end it at queue~$o$. Note that  
$\Lambda^k_i$ and~$M^k_o$ are known parameters for the algorithms, 
which can be obtained by the edge server leveraging the CAMs 
periodically transmitted by the vehicles.
Vehicles move from a generic queue $i$ to  queue $j$ according to
probability~$\alpha_{ij}$. 

Given the above model, our decision variables are the service
rates~$\mu_i$ of queues $q_i\in\Qc$, and the transition
probabilities~$\alpha_{i,j}$ 
(although some $\alpha_{ij}$ can be fixed, e.g., in \Fig{scenario},
$\alpha_{4,5}=1$). 
The total incoming flow for queue~$q_i$ is denoted with~$\lambda_i$
and depends on $\Lambda_i^k$, $M_o^k$, $\alpha_{i,j}$; it is given by: 
\begin{equation}
\label{eq:lambda-i-generic}
\lambda_i= \sum_{k\in\Kc}\Lambda^k_i
+\sum_{h\in\Qc}\alpha_{h,i}\left(\mu_h(1-\pi^0_h)-\sum_{k\in\Kc}M^k_h\right)
\end{equation}
where $\pi^0_i$ denotes the probability~that queue~$q_i$ is empty. 
Eq.\,\Eq{lambda-i-generic} can be read as follows: the flow entering queue~$q_i$ is equal to the flow of vehicles that begin their
journey therein, plus a fraction~$\alpha_{h,i}$ of the vehicles that
exit other queues~$q_h$, but do not leave yet the road topology.

\subsection{A path-based view} 
\label{sec:sub-paths}

Modeling road topologies as queue networks~\cite{rev1,rev1b,rev3-1,rev3-2} is a sensible methodology to estimate the travel times of different vehicles or group thereof. However, it is often cumbersome to make {\em decisions} based on queue models, owing to the large number of variables therein (the routing probabilities) and the complex way in which such probabilities influence the {\em distribution} of the travel times of vehicles of the same flow. 
To allow for quicker and higher-quality decisions, we now leverage the notion of {\em path}, i.e., a sequence of queues with a given source and destination, and per-flow path probabilities.

Vehicle with the same source and destination are said to belong to a {\em flow}  
 $k\in\Kc$, with $\Kc$~identifying the set of flows. Edge servers can obtain information 
on the vehicles' destination from the CAMs they send, or 
through application-layer communication. Indeed, it is fair to assume that vehicles 
benefiting from a service, e.g., navigation, agree to disclose their destination, 
provided that proper anonymization and privacy mechanisms~\cite{zhu2009security} are in place.

Vehicles of the same flow
may nonetheless follow different
trajectories, hence, traverse different sequences of queues.
We define such sequences of queues as {\em paths}~$w\in\Wc$.
Each path  $w$  is an array including the queues traversed
by vehicles taking it; given a queue~$q_i\in\Qc$, we  
write~$q_i\in w$ if  path~$w$ includes~$q_i$. 
The probability that vehicles of flow~$k$ take
path~$w$ is denoted by $p_w^k$; each path~$w$ is used by one flow~$k$ only, indicated
as~$\kappa(w)\in\Kc$. 
Introducing paths
allows for a flexible relationship between flows and
queues, thus enhancing the realism of our model, while keeping its
computational complexity low.
Indeed, the set~$\Wc$ and the $\kappa(w)\in\Kc$~values can be pre-computed, hence, 
they do not impact the complexity of the decision-making algorithm.

\begin{figure}[th!]
\centering
\includegraphics[width=0.85\columnwidth]{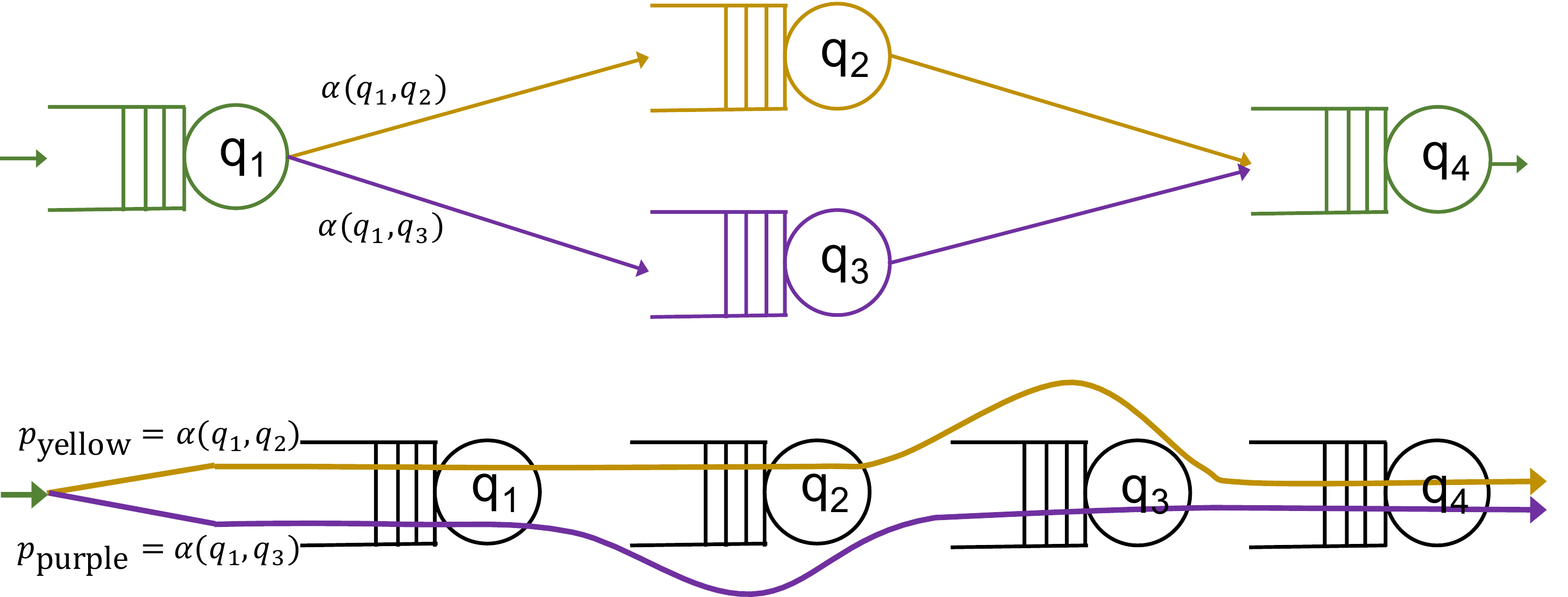}
\caption{Original queue network (top) and path view (bottom).
    \label{fig:equivalent}
    } %
\end{figure}

\noindent\rule{1\columnwidth}{0.5mm}
\begin{example}[Path-based view]
    \label{ex:equivalent}
Consider a single flow and the topology in \Fig{equivalent}(top), where vehicles
visit~$q_1$, then either $q_2$ or~$q_3$, and then~$q_4$. Since there
are two routes that the vehicles can follow, there are two paths in~$\Wc$, namely, $w_\text{yellow}=[q_1,q_2,q_4]$ and $w_\text{blue}=[q_1,q_3,q_4]$.
As for the probabilities~$p_w^1$ with which each path is taken, they
depend on the transition probability~$\alpha_{1,2}$
and~$\alpha_{1,3}$; specifically,~$p_\text{yellow}=\alpha_{1,2}$
and~$p_\text{purple}=\alpha_{1,3}$. The resulting parallel-path
view is displayed in \Fig{equivalent}(bottom).
\end{example}
\noindent\rule{1\columnwidth}{0.5mm}

Importantly, given  the edge-based policy, 
vehicles are assigned a path (according to the $p_w^k$~probabilities)
at the beginning of their travel across the considered road topology, 
before they enter the first queue. Thus, given the path it takes, the
queues traversed by a vehicle are unequivocally determined.

The probabilities~$p_w^k$ that vehicles of flow~$k$ take path~$w$
can be expressed as a function of the probabilities~$\alpha_{i,j}$ as:
\begin{equation}
\label{eq:pwk}
p_w^k=\begin{cases}
\prod_{n=2}^{|w|}\alpha_{w[n-1],w[n]} & \mbox{ if } \kappa(w)=k\\
0 & \text{otherwise,}
\end{cases}
\end{equation}
where~$w[n]\in\Qc$ is the $n$-th queue included in path~$w$.
Clearly, the $p_w^k$~probabilities have to sum up to one:
\begin{equation}
\label{eq:pwk-sum-one}
\sum_{w\in\Wc}p_w^k=1,\quad\forall k\in\Kc.
\end{equation}

We  stress that, while each path is unequivocally
associated with one flow, the opposite does not hold. 
With reference to \Fig{scenario}, flow~$1$ is
associated with one path~$[q_1,q_2,q_5]$ only, while 
flow~$2$ can use path~$[q_2,q_3,q_5]$
or~$[q_2,q_4,q_5]$. 
As for paths and queues, there is a many-to-many relationship between
them. In \Fig{scenario}, 
queues~$q_1$ and~$q_4$ belong to one path each, queues~$q_2$ and~$q_3$
to two paths each, and queue~$q_5$ to all three paths.

When considering paths, the local arrival rate at each queue $q_i$ 
in \Eq{lambda-i-generic} depends on: 
the paths $q_i$ belongs to, the probability that flows take
those paths, and the arrival rate of those flows. Specifically,
\begin{equation}
\label{eq:lambda-p}
\lambda_i=\sum_{k\in\Kc}\Lambda^k\sum_{w\in\Wc\colon q_i\in w}p^k_w.
\end{equation}
Notice how \Eq{lambda-p} expresses the same quantity as \Eq{lambda-i-generic} earlier. However, \Eq{lambda-p} leverages our path-based representation, while \Eq{lambda-i-generic} considers the elementary building blocks of the system, namely, queues and the routing probabilities between them.

\section{Flow Travel Times}
\label{sec:times}

In the following, we leverage the path-based view of the system to
characterize the travel time experienced by each flow~$k$. 
Let us first denote with $f_i(t)$ the probability density function (pdf) of the sojourn time at an
individual queue~$q_i\in\Qc$.  Such pdf is a function of $\lambda_i$ and
$\mu_i$, according to an expression that depends on the statistics of
the queue arrival process and the service time.
Then,  recalling that vehicles taking path~$w$ traverse all queues in $w$,  the travel time associated with path~$w$ is the sum of the sojourn times at all queues therein. The pdf of such a time is the $|w|$-way convolution of the individual pdfs associated with each queue, i.e.,
$f_w(t)=\textsf{Conv}_{q_i\in w}f_i(t)$.
By integrating $f_w(t)$, we compute the cumulative density functions
(CDFs) and take their 
Laplace transform, thus obtaining:
\begin{equation}
\label{eq:cdf-s-generic}
\mathbf{F}_w(s)=\frac{1}{s} \prod_{q_i\in w} \mathbf{f}_i(s),
\end{equation}
where $\mathbf{f}(s)=\mathcal{L}[f(t)](s)$ is the Laplace transform
of~$f(t)$. 

Anti-transforming \Eq{cdf-s-generic}, we can  obtain the CDF of the path-wise travel time:
$F_w(t)=\mathcal{L}^{-1}\left[\mathbf{F}_w(s)\right](t)$.
We recall that~$F_w(t)$ is a function of the control
variables~$\mu_i$ and~$p_w^k$ (i.e., $\alpha_{h,i}$), which appear in the pdf~$f_i(t)$. As
mentioned, the
actual form of~$f_w(t)$ and~$F_w(t)$, as well as of the above Laplace
transforms, depends on the queue arrival process and service time
distribution. In  many cases of
interest, $\mathbf{F}_w(s)$ can be expressed as a ratio
between polynomials, and anti-transformed into a summation of terms of
type~$At^n\mathrm{e}^{t\tau}$, as shown in \Sec{sub-special}. 

Next, we  move from paths to flows.
Intuitively, the travel time of flow~$k$ can be expressed as the sum of the travel times of 
paths~$w$'s that can be used by vehicles of flow~$k$, each weighted by probability~$p_w^k$. 
We first exploit 
$F_w(t)$ to write the
probability~$\delta_w(\hat{t})$ that the travel time of a vehicle
taking path~$w$ 
exceeds a value~$\hat{t}$:
\begin{equation}
\label{eq:prob-w}
\PP(\text{travel time on path $w$} > \hat{t})=1-F_w(t)|_{t=\hat{t}}
\triangleq \delta_w(\hat{t}).
\end{equation}
We then move to a
flow-wise equivalent of $\delta_w(\hat{t})$, 
combining the $\delta_w$ path-wise probabilities  with
the values~$p_w^k$, expressing the probability that a vehicle of flow~$k$ takes path~$w$, and write:
\begin{equation}
\label{eq:prob-k}
\PP(\text{travel time of  flow $k$} \mathord{>}
  \omega^k)= \sum_{w\in\Wc}p^k_w\delta_w(\omega^k) 
 \triangleq  \delta^k(\omega^k) \,.
\end{equation}
In \Eq{prob-k}, $\omega^k$ is the flow-wise target travel time, 
e.g., the ratio of the distance between the flow source and
destination to the desired average speed between them. 

It is worth stressing that, if closed-form expressions of~$F_w(t)$ 
do not exist or are too complex to manage, the steps above can still be performed numerically, obtaining the same results. 
In that case, however, we do not resort to Laplace transforms, 
and instead directly compute the required convolutions and integrals.

\subsection{M/M/1 queues and batch arrivals}
\label{sec:sub-special}

For sake of concreteness and as a useful example, in the following
we show how the above expressions particularize to
the relevant
cases~\cite{heidemann1996queueing,van2006empirical,zhou2018begin,ning2019vehicular,zhao2018joint,rev3-1}
where the road segments are modeled as $M/M/1$ or $M^X/M/1$ 
queues~\cite{carlo-secon12,mukhopadhyay2017approximate}, 
with $X$ denoting  the (discrete) distribution of the vehicle batch size. 

Let us start from the $M/M/1$ case and consider the per-path travel
times presented in
\Sec{times}. 
Given the expression of the pdf of the sojourn time at an $M/M/1$
queue~$q_i$, i.e.,  
$f_i(t)=(\mu_i-\lambda_i)\mathrm{e}^{-(\mu_i-\lambda_i)t}u(t)$~\cite{mm1times},
where~$u(t)$ is the step function,
we can write:
$\mathbf{f}_w(s)=\prod_{q_i\in
  w}\frac{\mu_i-\lambda_i}{s+\mu_i-\lambda_i}$. 
Thus, the CDF is given by:
$\mathbf{F}_w(s)=\frac{1}{s}\prod_{q_i\in
  w}\frac{\mu_i-\lambda_i}{s+\mu_i-\lambda_i}$, which 
is the ratio between two polynomials in~$s$: 
a numerator of degree~0 and a denominator of
degree~$|w|+1$. It has one pole in~$s=0$ and additional~$|w|$ poles,
one for each queue~$q_i\in w$, at~$s=\mu_i-\lambda_i$. It is well
known that this expression can be decomposed into partial fractions~\cite{kung1977fast}:
\begin{equation}
\label{eq:trasf-cdf-frac}
\mathbf{F}_w(s)=\frac{A_{0,w}}{s}+\sum_{q_i\in
  w}\frac{A_{i,w}}{s+\mu_i-\lambda_i} 
\end{equation}
where 
\begin{equation}
\label{eq:a-i}
A_{i,w}=-\prod_{q_j\in w\colon j\neq i}\frac{\mu_j-\lambda_j}{(\mu_j-\lambda_j)-(\mu_i-\lambda_i)},
\end{equation}
and $A_{0,w}=1$.
Anti-transforming \Eq{trasf-cdf-frac}, we can write the CDF of the
travel time on path~$w$ as:
\begin{equation}
\label{eq:cdf-sojourn-w}
F_w(t)=u(t)\left(1+\sum_{q_i\in w}A_{i,w} \mathrm{e}^{-(\mu_i-\lambda_i)t}\right).
\end{equation}

Therefore, combining \Eq{prob-w}, \Eq{prob-k}, and \Eq{cdf-sojourn-w},
and exploiting \Eq{pwk-sum-one}, $\delta^k(\omega^k)$ can be
written as:
\begin{equation}
\label{eq:deltak-mm1}
\delta^k(\omega^k)
=-\sum_{w\in\Wc}\sum_{q\in w}p_w^k A_{i,w}
    \mathrm{e}^{-(\mu_i-\lambda_i)\omega^k}. 
\end{equation}

In the $M^X/M/1$ case, vehicles arrive in batches,
with size characterized by a given probability mass function (pmf).
Let the pmf be $g_n=\PP(\text{batch size}=n)$, $1\leq
n \leq G$, with $G$ being the maximum batch size. The expected batch size is $\bar{g}=\sum_{n=1}^G ng_n$.
Then the Laplace transform of the path travel time is~\cite[Sec.~4.2.1]{medhi2002stochastic}:
\begin{equation}
\label{eq:mxm1-pdf}
\mathbf{F}_w(s)=\frac{1}{s} \prod_{q_i\in w} \frac{(1-\rho_i)(1-s)}{\mu_i(1-s)-\lambda_i s(1-\Gamma(s))},
\end{equation}
where~$\rho_i=\frac{\lambda_i \bar{g}}{\mu_i}$ is the queue utilization and $\Gamma(s)=\sum_{n=1}^G g_ns^n$ is the probability generation function  of the batch size pmf.
It is important to highlight that \Eq{mxm1-pdf} is a ratio of
polynomials in~$s$: a
numerator of degree~$|w|$ and a denominator of
degree~$1+|w|(G+1)$. Hence, using the same methodology as above, \Eq{mxm1-pdf}
can be decomposed into partial
fractions and anti-transformed, yielding an expression of~$\delta^k(\omega^k)$ similar to \Eq{deltak-mm1}.

\section{Problem formulation}
\label{sec:problem}

Our high-level goal is 
to keep the fraction of vehicles of each flow~$k$, whose travel time
exceeds the target~$\omega^k$, as low as possible. 
To ensure fairness among vehicles of different flows, we
formulate such a goal as the following min-max objective:
\begin{equation}
\label{eq:obj-generic}
\min_{p^w_k,\mu_i}\max_{k\in\Kc}\sum_{w\in\Wc}p^k_w\delta_w(\omega^k)\,.
\end{equation}

The optimization variables are the probabilities~$p^w_k$ (hence the
$\alpha_{i,j}$) and the
service rates~$\mu_i$, 
which, as also exemplified in \Sec{sub-special}, appear in
the objective function in \Eq{obj-generic} via the CDF in 
\Eq{cdf-s-generic}. 
Ingress and egress rates~$\Lambda_i^k$ and~$M_i^k$ are input
parameters, as are the maximum service rates~$\mu^{\max}_i$ and the
target flow travel times~$\omega^k$. The per-queue arrival rates~$\lambda_i$
are auxiliary variables, 
whose dependency on the decision variables is specified in 
\Eq{lambda-p}.
Furthermore, we need to impose the constraints in \Eq{mu-max}  and \Eq{pwk-sum-one}.

{\bf Reconstructing the $\alpha$-values.}  
Given the optimal~$p_w^k$ values, the~$\alpha_{i,j}$ variables can be
recovered by solving a system of equations of the same type as \Eq{pwk}, where
the $\alpha_{i,j}$-values are the unknown and the $p_w^k$-values are
given. The system can be linearized by taking the logarithm of all
variables, i.e., re-writing \Eq{pwk} as:  
$\log p_w^k=\sum_{n=2}^{|w|}\log\alpha_{w[n-1],w[n]}$, $\forall w\in\Wc$.

The fact that the system has a unique solution is ensured by the 
proposition below; the intuition behind it is that the number
of paths grows faster than the number of junctions, hence, there are
more equations than variables.  It follows that, if it
  exists, the solution  is unique. 
\begin{proposition}
\label{prop:npaths}
In any topology, the number of paths is strictly higher than the number of junctions.
\end{proposition}
\begin{IEEEproof}
Let us start with a single path, hence, zero junctions. Then junctions
are added one at a time, starting from queues~$q_i$ that are already part of at least one path; by so doing, the number of junctions increases by one. 
Also, a new path is created for every path that included~$q_i$, hence, the number of paths grows by {\em at least} one.
\end{IEEEproof}

\subsection{Problem complexity and solution approaches}

As the road topology may include several lanes and road
stretches, it is important to understand the complexity of our
problem, as well as the viable solution approaches that can be pursued.
Indeed, such a  complexity  has an impact on the
computational resources required  at the edge to provide timely
policy updates. 

We first look at the expression for~$F_w(t)$ in \Eq{obj-generic}, which can
be fairly complex even for simple queuing systems, as exemplified in
\Sec{sub-special} for the $M/M/1$ case. In particular, the
$A_{i,w}$~coefficients multiplying the exponential terms can have any sign. 
One may be tempted to conclude that nothing can be said about the
convexity of the problem; instead, it is possible to prove that:
    (i) in the most general case, the expression is monotonic, namely, non-decreasing;
    (ii) in a wide set of relevant cases, it is convex.

\begin{theorem}
\label{thm:monotonic}
Optimizing \Eq{obj-generic} subject to constraints
\Eq{mu-max}, \Eq{pwk-sum-one}, and \Eq{lambda-p}
requires solving a monotonic optimization problem.
\end{theorem}\begin{IEEEproof}
The monotonicity of the objective function \Eq{obj-generic} derives from the fact that the
quantity~$\delta_w(\hat{t})$ can be obtained according to \Eq{prob-w},
by computing a CDF in a specific value~$\hat{t}$. CDFs are
non-decreasing functions and
the~$p^k_w$ coefficients are probabilities, hence, non negative. It follows that the quantity within the summation of \Eq{obj-generic} is a conic combination of non-decreasing expressions, which is itself non-decreasing. Similarly, taking the (point-wise) maximum over all flows~$k$ preserves monotonicity; therefore, \Eq{obj-generic} as a whole is monotonic.
As for the constraints,  \Eq{pwk-sum-one} and \Eq{lambda-p} are linear, hence, they meet the requirement that equality constraints of monotonic optimization problems are affine.
\end{IEEEproof}
Monotonic optimization problems~\cite{tuy2000monotonic} can be solved by mapping them into a set of 
convex sub-problems. Individual sub-problems are then solved to optimality in polynomial time, 
while the original (monotonic) problem can be solved arbitrarily closed to the optimum, 
at the cost of increasing the number of sub-problems. 

Thm.\,\ref{thm:monotonic} does not rely on any
specific travel time distribution, but only on it being characterized by  a
CDF. Indeed, Thm.\,\ref{thm:monotonic} still holds for service time
distributions that do not result in a closed-form expression of the
sojourn time CDF, as, e.g., in the case of $M/D/1$ queues. The wide scope of Thm.\,\ref{thm:monotonic} reflects the
generality of our framework, which {\em does not depend on any
  restrictive modeling assumption} on the queue service time,
but works unmodified under any service time distribution.

We can prove an additional result showing that, under mild conditions, 
solving the problem of optimizing \Eq{obj-generic} subject to constraints \Eq{mu-max}, 
\Eq{pwk-sum-one}, and \Eq{lambda-p} can be reduced to a convex problem. 
The key mathematical concept we leverage is {\em log-convexity}. 
Simply put, a function is log-convex if its logarithm is convex~\cite{kingman1961convexity}. 
As an example, the service time distribution of an M/M/1 queue, for $t\geq 0$,
is~$f_i(t)=(\mu_i-\lambda_i)\mathrm{e}^{-(\mu_i-\lambda_i)t}$ and its logarithm 
is~$\log(\mu_i-\lambda_i)+(\lambda_i-\mu_i)t$, which is linear in~$t$, hence, convex; 
it follows that~$f_i(t)$ is log-convex. Note that log-convexity is a more restrictive 
condition than convexity, hence, log-convexity implies convexity.

If all queues in the system have a log-convex sojourn time pdf, then the following important result holds.
\begin{theorem}
\label{thm:logconvex}
If all queues in the system have log-convex sojourn time pdfs, then the problem of
optimizing \Eq{obj-generic} subject to constraints
\Eq{mu-max}, \Eq{pwk-sum-one}, and \Eq{lambda-p}
reduces to a convex optimization problem.
\end{theorem}\begin{IEEEproof}
As shown in \Sec{times}, computing the distribution of per-path travel times involves 
convolutions and integrations of the sojourn time pdfs of the individual queues. 
Both these operations preserve the log-convexity property \cite{kingman1961convexity}; it follows that, 
if the hypothesis holds,  the per-path travel time CDFs are log convex, hence, convex.

Now, the objective function \Eq{obj-generic} is computed as per \Eq{prob-w}, 
by considering each CDF at a specific point~$\hat{t}$. Furthermore, the coefficients~$p_w^k$ 
are probabilities, hence, non negative. It follows that the quantity within the summation in 
\Eq{obj-generic} is a conic combination of convex expressions, which is itself convex. 
Similarly, taking the (point-wise) maximum over all flows~$k$ preserves convexity; 
therefore, \Eq{obj-generic} as a whole is convex. 
As for the constraints, \Eq{pwk-sum-one} and \Eq{lambda-p} are linear, hence, 
they meet the requirement that equality constraints of convex optimization problems are affine.
\end{IEEEproof}

The condition of Thm.\,\ref{thm:logconvex} is met by a very large number of relevant queuing systems, 
including the very popular~$M/M/1$~\cite{heidemann1996queueing,zhou2018begin,ning2019vehicular,zhao2018joint} and~$M^X/M/1$~\cite{alfa1995modelling} ones.

\subsection{Discussion and useful insights}

Thm.\,\ref{thm:monotonic} and Thm.\,\ref{thm:logconvex} imply that our problem can be efficiently solved by
commercial, off-the-shelf solvers like CPLEX or
Gurobi, especially when the conditions of Thm.\,\ref{thm:logconvex} hold and the problem is convex.
However, numerical solvers encounter significant numerical
difficulties when dealing with the objective \Eq{obj-generic}. As
demonstrated for instance in \Sec{sub-special} for the $M/M/1$ case, the~$A_{i,w}$ coefficients therein usually
contain a ratio of products of $\mu_i-\lambda_i$~terms: a small variation in any of the terms can change the sign of the whole coefficient and/or significantly alter its (absolute) value. The issue is compounded by the fact that such values, which can be very large, are then multiplied in \Eq{prob-k} by probabilities~$p_w^k$, which could be very small and/or be varied by very small quantities. State-of-the-art interior-points methods are engineered to deal with both challenges, however, their convergence can be slowed down -- or prevented altogether -- by  such numerical difficulties.

Gradient-based methods like the BFGS algorithm~\cite{bfgs} may 
perform better, and more reliably (albeit more slowly) 
converge.  
The basic intuition behind gradient-based methods
is to start from a feasible solution, and then improve it by 
performing, at every iteration, the change 
that results in the largest improvement of the objective function. 
However, we
can achieve faster convergence through an {\em ad hoc} algorithm
exploiting problem-specific knowledge.

To design an efficient solution, we look at the  case of
$M/M/1$ queues with the aim to  derive some insights of general validity. In
particular, an aspect worth investigating is the intuitive notion
that  equilibrium across queues and paths can lead to better
performance. To show this and pick a suitable metric to define such equilibrium
conditions, we  make the observation below.
\begin{theorem}
\label{thm:allequal}
Consider the quantity $\delta^k(\omega^k)$ in \Eq{cdf-sojourn-w} for a
fixed $t$, and assume that
$\mu_i-\lambda_i=d$, $\forall q_i\in w$.
Then transferring an amount $\epsilon>0$ of traffic from any
queue~$l\in w$ to any other queue~$m\in w$  always increases
$\delta_w(\omega^k)$, hence the value of the objective
\Eq{obj-generic}. 
\end{theorem}
\begin{IEEEproof}
Let us consider a tagged path $w$ and two additional paths $w'$ and
$w''$. Without loss of generality, let us move an amount of traffic,
$\epsilon$, from  queue $q_h \in w,\,w'$ to  queue $q_l \in w$, and
the same amount of traffic from $q_m \in w$ to $q_r \in w,\,w''$. 
Now, note that when we had $\mu_l-\lambda_l=\mu_m-\lambda_m= d$, the
transform of the sojourn time CDF \Eq{trasf-cdf-frac} had a double pole 
in~$d$ and,
according to the same partial fraction decomposition rules used to compute the $A_i$~coefficients 
in \Eq{a-i}, the path travel time distribution (see \Eq{cdf-sojourn-w}) reduces to:
$F_w^\text{old}(t)=1-\mathrm{e}^{-dt}-dt\mathrm{e}^{-dt}=1-\mathrm{e}^{-dt}(dt+1)$. 
Moving $\epsilon$~traffic from $q_l$ to a $q_m$ means replacing the two poles in~$d$ with two single poles, one in~$d+\epsilon$ and one in~$d-\epsilon$. \Eq{cdf-sojourn-w} then becomes:
$F_w^\text{new}(t)=1-\mathrm{e}^{-dt} [\frac{d+\epsilon}{2\epsilon}\mathrm{e}^{\epsilon t}-\frac{d-\epsilon}{2\epsilon}\mathrm{e}^{-\epsilon t}]$.

To prove the thesis, it must be~$F_w^\text{new}(t)\leq F_w^\text{old}(t)$, i.e., $F_w^\text{new}(t)-F_w^\text{old}(t)\leq 0$. Considering only the expressions multiplied by~$-\mathrm{e}^{-dt}$, this is equivalent to imposing:
\begin{equation}
\label{eq:diff}
\frac{d+\epsilon}{2\epsilon}\mathrm{e}^{\epsilon t}-\frac{d-\epsilon}{2\epsilon}\mathrm{e}^{-\epsilon t}-dt-1\geq 0.
\end{equation}
We remark that the first member of \Eq{diff} tends to~$\infty$ when~$t\to\infty$, and to~$0$ when $t\to 0$. To assess the behavior of \Eq{diff} over the rest of its domain, we compute its derivative over~$t$, obtaining:
$\frac{d+\epsilon}{2}\mathrm{e}^{\epsilon
  t}+\frac{d-\epsilon}{2}\mathrm{e}^{-\epsilon t}-d$, which 
must be be greater than zero. The derivative can be re-written as:
$ \frac{d}{2}\left(\mathrm{e}^{\epsilon t}+\mathrm{e}^{-\epsilon
    t}\right)+\frac{\epsilon}{2}\left(\mathrm{e}^{\epsilon
    t}-\mathrm{e}^{-\epsilon t}\right)-d$. 
The second member is clearly non negative, since~$\mathrm{e}^{\epsilon
  t}\geq\mathrm{e}^{-\epsilon t}$ for any non-negative values of~$t$
and~$\epsilon$. The quantity~$\mathrm{e}^{\epsilon
  t}+\mathrm{e}^{-\epsilon t}$ is equal or greater than~$2$ --
specifically, it has a minimum at~$2$ when~$t=0$ --, hence, the first
member of the derivative is always greater or equal to~$d$. 
In conclusion, the derivative of the quantity at the first member of
\Eq{diff} is always non negative, hence, the inequality \Eq{diff} is
valid, hence, moving~$\epsilon$ traffic from a queue to another yields
a value of the CDF of the path travel time, $F_w^\text{new}(t)$, that
is lower than $F_w^\text{old}(t)$, for any $t$. Hence, the travel time increases, which proves the thesis.
\end{IEEEproof}

We can also prove the
following corollary, which holds in the case of multiple paths.
\begin{corollary}
\label{cor:multipath}
Let $k$ be a flow that can be routed through~$|\Wc|\geq 2$ paths, each
containing the same number~$|w|$ of queues, and assume that, for each
path $w$, $\mu_i-\lambda_i=d$ for any $q_i\in w$.  Then the lowest
value of the objective function in \Eq{obj-generic}  is attained when~$p_w^k=\frac{1}{|\Wc|},\forall w$.
\end{corollary}
\begin{IEEEproof}
The proof follows from the definition of~$\delta^k(\omega^k)$ given in
\Eq{prob-k}, which is a weighted sum of terms~$F_w(\omega^k)$. Let us
consider $w$ and $w'$ such that $\kappa(w)=\kappa(w')=k$. Moving a
fraction~$\delta$ of traffic from path~$w$ to  path~$w'$ means
increasing the weight corresponding to the $w$-term, and decreasing
the weight corresponding to the $w'$-term. At the same time,
considering that $F_w(t)$~functions are CDFs and CDFs are
non-decreasing, it also means increasing the $\delta^k(w)$ term and
decreasing the $\delta^k(w')$-term, which results in a larger value
for~$\delta^k(\omega^k)$, hence, for the objective function in \Eq{obj-generic}.
\end{IEEEproof}

In summary, we observed first that  equilibrium conditions are indeed associated with better performance. Second, the best metric to use when defining equilibrium is the difference between the service and arrival rate at each queue, i.e.,~$\mu_i-\lambda_i$. We leverage both insights while defining our algorithm, as explained next.

\section{The BH algorithm}
\label{sec:algo}

We now propose an iterative algorithm to solve the above optimization problem, and discuss its properties and novelty. We underline that the algorithm works unmodified for
a very wide range of queuing models, with different arrival processes, service time distributions, and number of servers.

\subsection{Goal and approach}

Thm.\,\ref{thm:monotonic} and Thm.\,\ref{thm:logconvex} imply that our problem can be solved by performing one or more {\em convex} optimizations, even when its objective does not have a closed-form expression. A prominent family of algorithms that solve convex problems is represented by {\em gradient-descent algorithms}
-- from the venerable Newton algorithm to more modern 
alternatives such as BFGS~\cite{bfgs} and stochastic gradient descent~\cite{sgd} (SGD), 
the latter widely used in machine-learning applications.

Gradient-descent algorithms work iteratively, by changing at each iteration one variable 
in the current solution, and effecting the change yielding the highest improvement 
in the objective function. If, as in our case, the problem is convex, {\em any} 
gradient-descent algorithm is guaranteed to converge to the optimal solution.

However, general-purpose algorithms such as BFGS or SGD are unaware of the underlying structure of the problem, and make their decisions in terms of individual variables, i.e., per-path probabilities. This may lead them to make changes that, due to their impact on other paths and flows, have to be undone at a later iteration; the final result is that the optimal is reached in a longer time than needed.

To avoid this problem, we design an iterative algorithm called {\em bottleneck-hunting} (BH). Our high-level design objectives can be summarized as follows:
   (i) follow the same {\em philosophy} as gradient-descent, i.e., iteratively improving a solution;
    (ii) make {\em higher-level} decisions, accounting for flows and paths instead of individual optimization variables;
    (iii) avoid making decisions that need to be undone at later stages.
The resulting algorithm is described in \Sec{sub-descr} and analyzed in \Sec{sub-props}. 
It has the same worst-case performance and convergence properties of gradient-based alternatives but, 
thanks to its awareness of problem-specific information, it exhibits faster {\em average-case} 
performance.

\subsection{Algorithm description}
\label{sec:sub-descr}

As discussed above, the BH algorithm 
finds a solution to the problem specified in \Sec{problem} faster than general-purpose, gradient-descent-based alternatives.
Such a faster convergence is achieved by leveraging problem-specific knowledge and information in order to reduce the number of solutions to try out, hence, of algorithm iterations.
Specifically, Thm.\,\ref{thm:allequal} and \Cor{multipath} show that,  if a
solution where all $M/M/1$ queues of each path have the same load is
feasible, then the solution is optimal. It follows that, in those
cases, it is never necessary to {\em widen} the gap between the most-
and least-loaded queues of any path. We take this as a guideline to
design the BH algorithm, which iteratively improves a current solution
similarly to a gradient-based approach, but avoiding widening the
aforementioned gap whenever possible.

With reference to \Alg{bh}, at every iteration, BH moves a
fraction~$\phi$ of flow~$k^\star$'s traffic from path~$w^\star$ to
path~$w'$. The fraction~$\phi$ changes across iterations, and is
initialized (\Line{init-phi}) to a value~$\phi_0$. Then, at every
iteration, BH identifies (\Line{cq}) a set \path{CQ} of {\em critical
  queues}, that is, queues that: (i) belong to two paths~$w_1$
and~$w_2$; (ii) are not the most loaded queues in~$w_1$; and, (iii)
 increasing their load by a fraction~$\phi$ of the
 traffic~$\Lambda^{\kappa(w_2)}$ of flow~$\kappa(w_2)$ renders such queues  
 the most loaded ones in~$w_1$.
It follows that the algorithm will try to avoid routing additional traffic 
on critical queues in \path{CQ} if possible. Based on \path{CQ}, a set \path{CP} of 
{\em critical paths}, i.e., paths containing at least one critical queue, is identified in \Line{cp}.

Next, BH identifies the set \path{FA} of flows that can be acted
upon. Flows using at least one non-critical path are tried first
(\Line{fa-ncp}); if no such flow exists (\Line{fa-check}), then
\path{FA} is extended to include all flows in~$\Kc$
(\Line{fa-all}). The flow~$k^\star$ to act upon is chosen in
\Line{kstar}: considering the min-max nature of objective
\Eq{obj-generic}, $k^\star$~is the flow for which the summation in
\Eq{obj-generic} is largest. For the same reason, the path~$w^\star$
to {\em remove} vehicles from is chosen as the most loaded one, hence,
the one associated with the largest term in \Eq{obj-generic}. 
Similarly, the path~$w'$ to {\em add} traffic to is selected (\Line{wprime}) 
as the least-loaded one among those used by~$k^\star$; 
note that this implies choosing a non-critical path if such paths exist.

\begin{algorithm}[t]
\caption{The bottleneck-hunting (BH) algorithm\label{alg:bh}}
\begin{algorithmic}[1]

\State{$\phi\gets\phi_0$} \label{line:init-phi}
\While{$\textbf{true}$}
\State{$\texttt{CQ}\gets\{q_i\in\Qc\colon\exists w_1,w_2\in\Wc\colon
  q_i\in w_1\wedge q_i\in w_2 \wedge (\mu_i-\lambda_i) -\min_{q_j\neq q_i\in w_1} (\mu_j-\lambda_j) \leq\phi\Lambda^{\kappa(w_2)}\}$} \label{line:cq}

\State{$\texttt{CP}\gets\{w\in\Wc\colon w\cap\texttt{CQ}\not\equiv\emptyset\}$} \label{line:cp}
\State{$\texttt{FA}\gets\{k\in\Kc\colon\exists w\in\Wc\setminus\texttt{CP}\colon \kappa(w)=k\}$} \label{line:fa-ncp}
\If{$\texttt{FA}\equiv\emptyset$} \label{line:fa-check}
\State{$\texttt{FA}\gets\Kc$} \label{line:fa-all}
\EndIf

\State{$k^\star\gets\arg\max_{k\in\texttt{FA}} \delta^k(\omega^k)$} \label{line:kstar}
\State{$w^\star\gets\arg\max_{w\in\Wc\colon \kappa(w)=k^\star} \delta_w(\omega^{k^\star})$} \label{line:wstar}
\State{$w'\gets\arg\min_{w\in\Wc\colon\kappa(w)=k^\star} \max_{q_i\in w}(\mu_i-\lambda_i)$} \label{line:wprime}

\If{$\textsf{does\_improve}(k^\star,w^\star,w',\phi)$} \label{line:improves}
\State{$p_{w^\star}^{k^\star}\gets p_{w^\star}^{k^\star}-\phi$} \label{line:pminus}
\State{$p_{w'}^{k^\star}\gets p_{w'}^{k^\star}+\phi$} \label{line:pplus}
\Else
\State{$\phi\gets\frac{\phi}{2}$} \label{line:halfphi}
\If{$\phi<\phi^{\min}$} \label{line:phimax}
\State{$\textbf{return}$} \label{line:break}
\EndIf
\EndIf

\EndWhile
\end{algorithmic}
\end{algorithm}

In \Line{improves}, BH calls the function {\sf does\_improve}, which
recomputes \Eq{obj-generic} and checks whether it improves by moving a
fraction~$\phi$ of flow $k^\star$'s vehicles from path~$w^\star$ to
path~$w'$. 
Indeed, the objective may not improve if the current value of~$\phi$
is too high, i.e.,  moving  a fraction of
traffic $\phi$ increases the traffic intensity on $w'$ too much. In this case, such 
action should be performed at a later iteration, when $\phi$ will be
smaller (\Line{halfphi}). If the objective improves, 
the $p_w^k$~variables are updated accordingly 
(\Line{pminus}--\Line{pplus}). 
The algorithm terminates when $\phi$~drops below the minimum value~$\phi^{\min}$ (\Line{phimax}).

\subsection{Analysis and discussion}
\label{sec:sub-props}

We now analyze two main aspects of BH: (i) its computational complexity, and (ii) its optimality 
in relevant practical cases. 
About the former, we prove that BH exhibits a remarkably low, namely, linear complexity.
\begin{proposition}
\label{prop:complex}
The BH algorithm has linear worst-case computational complexity.
\end{proposition}\begin{IEEEproof}
At every iteration, \Alg{bh} identifies the most congested path~$w^\star$ of 
the most congested flow~$k^\star$, and moves a fraction~$\phi$ away from it. 
This happens for at 
most~$\left\lceil\frac{1}{\phi}\right\rceil\leq\left\lceil\frac{1}{\phi^{\min}}\right\rceil$ times; 
after that, there would be no more traffic routed on~$w^\star$.

The whole process is repeated for at most~$\left\lceil\log_2\frac{\phi_0}{\phi^{\min}}\right\rceil$ different values of~$\phi$, giving a total complexity of:
\begin{equation}
\label{eq:complexity}
\left\lceil\log_2\frac{\phi_0}{\phi^{\min}}\right\rceil\left\lceil\frac{1}{\phi^{\min}}\right\rceil|\Kc||\Wc|=O(|\Kc||\Wc|)\,.
\end{equation}
\end{IEEEproof}

Furthermore, if the objective function is convex (i.e., if Thm.\,\ref{thm:logconvex} holds), then BH is guaranteed to converge to the optimal solution:
\begin{proposition}
\label{prop:opti}
If the objective function \Eq{obj-generic} is convex, then the BH algorithm converges to the optimum.
\end{proposition}
\begin{IEEEproof} 
To begin with, we observe that the only case in which
BH and Newton's algorithm may behave differently is when the test in
\Line{fa-check} returns {\bf false}, i.e., when there is at least one
non-critical path. In this case, BH only increases the traffic on
non-critical paths, while Newton's algorithm could increase the
traffic of any path.
Increasing the traffic of a non-critical path is never detrimental to
the objective \Eq{obj-generic}, as traffic is moved from a more-loaded path
(as per \Line{wstar}) to a less-loaded one. 

Repeating such a process will eventually make the path~$w'$ critical;
the algorithm will then move on to other non-critical paths and flows
in \path{FA}, arriving to a situation where 
all paths are critical. 
At this point, two cases are possible: 
either the current solution is optimal, and no changes will be made
until~$\phi$ drops below~$\phi^{\min}$, or further improvements are
possible, in which case BH will behave like Newton's algorithm for the
remaining iterations, and reach the optimal solution. 
\end{IEEEproof}

\begin{figure}
\centering
\includegraphics[width=.48\columnwidth]{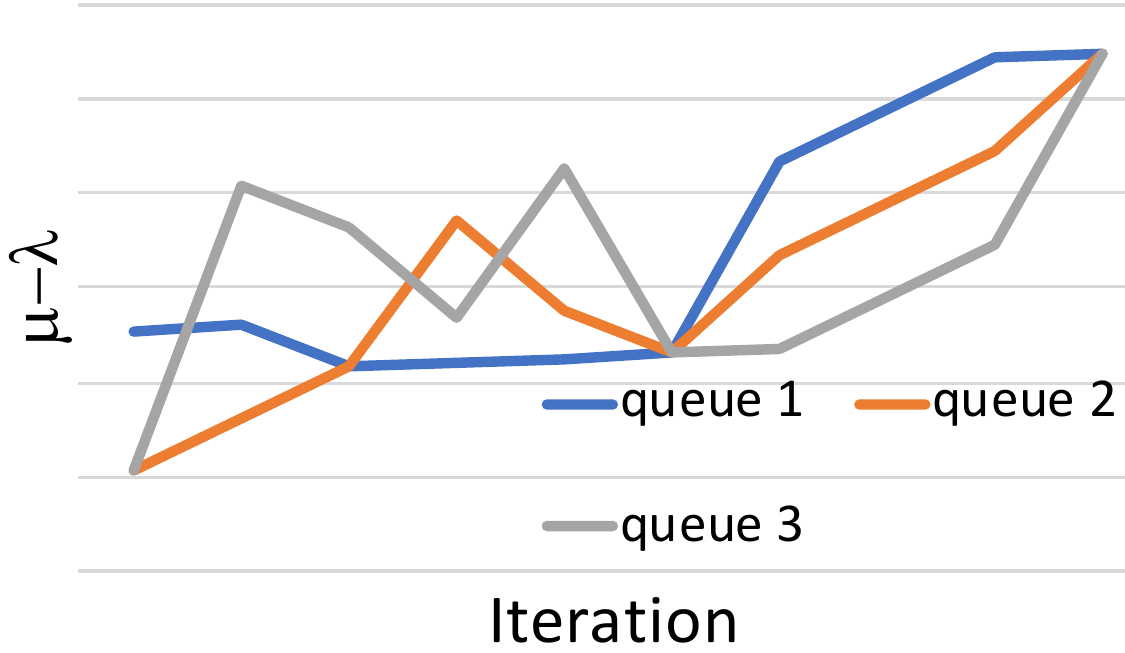}
\includegraphics[width=.48\columnwidth]{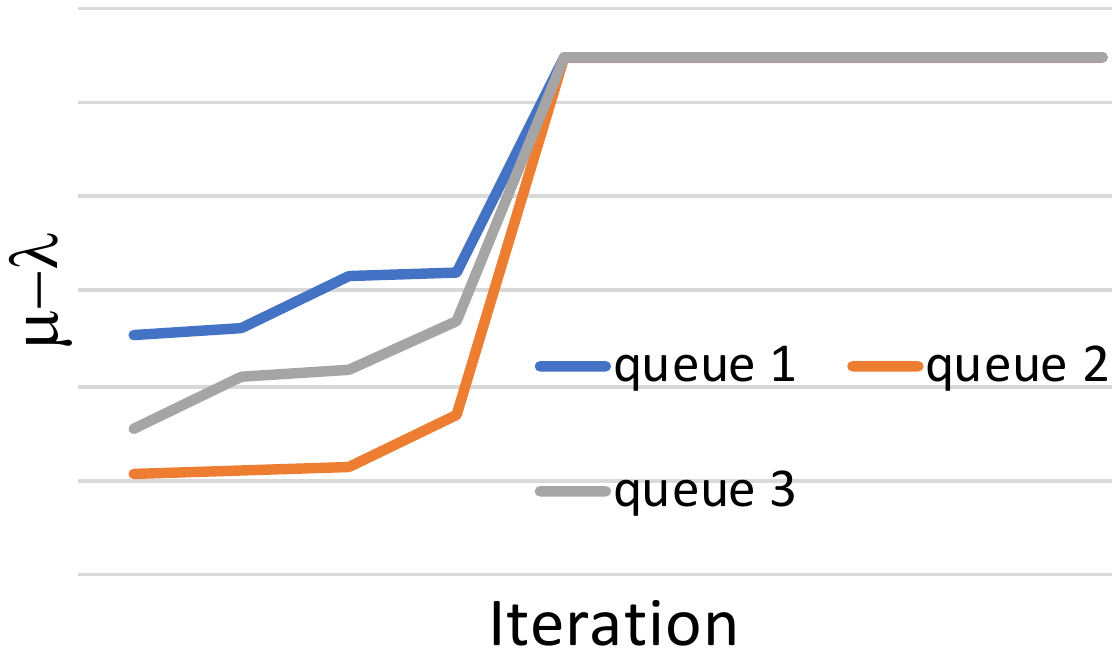}
\caption{Example of the evolution of the~$(\mu_i-\lambda_i)$ quantity in a toy scenario, for Newton's algorithm (left) and BH (right).
    \label{fig:newton-n-us}
} %
\end{figure}

The intuition behind \Prop{opti} is that BH 
does not waste iterations making changes that would eventually have to
be reversed. Even if the {\em worst-case} complexity proven in
\Prop{complex} is the same as other algorithms, the {\em actual}
number of iterations needed to converge is much smaller (see \Sec{convergence}). 
The difference between BH and gradient-based algorithms is exemplified
in \Fig{newton-n-us}, in the case of a path with three $M/M/1$
queues. Both Newton's algorithm (left) and BH (right) converge to the
same optimum solution; also, in both cases the {\em lowest}
$(\mu_i-\lambda_i)$ value  increases from one iteration to the
next. However, Newton's algorithm could occasionally decrease the
value $(\mu_i-\lambda_i)$  for some of the least busy queues, and then
be forced to undo those changes. BH, instead, never does so,
and converges to the optimum faster.

Notice that, \Prop{opti} is only guaranteed to hold under the same conditions 
as for Thm.\,\ref{thm:logconvex}, i.e., when the sojourn time pdfs are log-convex. 
However, as highlighted by our performance evaluation in \Sec{results}, 
BH exhibits the same desirable behavior under a much wider range of conditions, 
including those when no closed-form expression of sojourn time distributions exists, e.g., for $M/D/1$~queues.

Finally, we remark that BH exhibits fundamental differences with
respect to multi-path packet routing. First, routing protocols (including
state-of-the-art QUIC implementations~\cite{de2017multipath}) make
their decisions based on the {\em average} latency of each route, as
opposed to the full latency distribution. Second, routing protocols
look at either end-to-end or next-hop latencies, 
while BH accounts for individual road segments and how each of them contributes to the travel times.

\section{Validation Methodology}
\label{sec:validation}

\begin{figure}
\centering
\includegraphics[width=0.9\columnwidth]{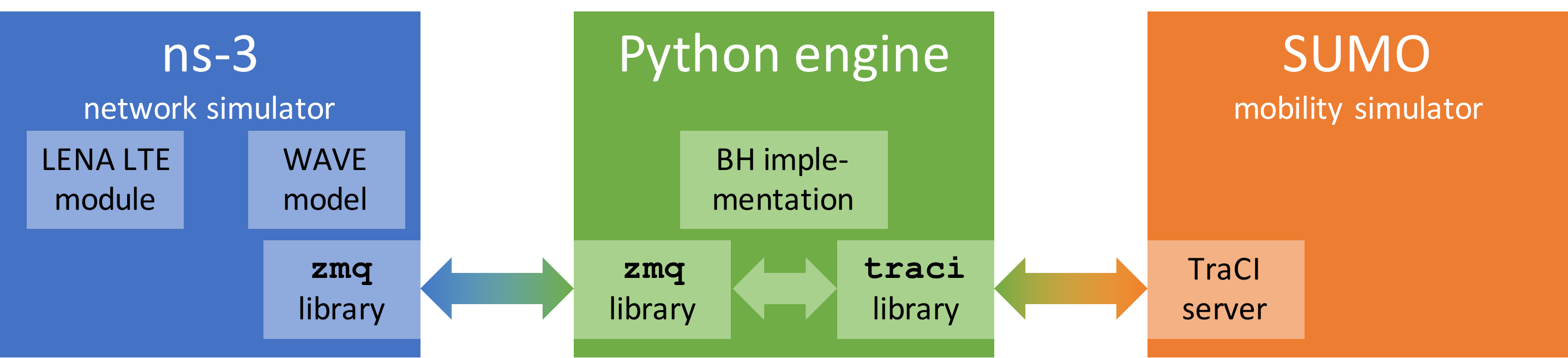}
\caption{Our validation framework, integrating the ns-3 network simulator,
a Python engine including the BH implementation, and the SUMO mobility simulator. %
\label{fig:validation}
} %
\end{figure}

 To assess its performance, we integrate BH within a complete validation
environment, as represented in
\Fig{validation}. Policies summarized by
the~$p_w^k$ and $\mu_i$ variables are defined by BH, which is implemented within a 
Python engine. Such decisions are then relayed to the ns-3 network simulator,
which is in charge of simulating:
   (i)  the network traffic generated by vehicles and by the infrastructure, and (ii) the actualizers, which, based on
   the most recent CAMs,  implement 
   the edge-defined policy.  
 Specifically, for the navigation service, 
    vehicles of flow~$k$ are randomly assigned a path~$w$ following the~$p_w^k$ 
    probabilities and are instructed accordingly through DENMs. 
    For the lane-change service, within every time window (e.g., 30 seconds), 
    the vehicles of each flow~$k$  having more free space in the destination lane (detected as per~[6])  
    are instructed through DENMs to  change lane, so as to honor probabilities $p_w^k$'s. 
In the case of the lane-change service, upon receiving a DENM, vehicles engage their neighbors in a
communication following the  protocol in \cite{autonet30-commag}, to
coordinate their  manoeuvres and avoid collisions. 
In both  the lane-change and the 
navigation cases, mobility is simulated via
SUMO %
and vehicles traversing stretch of road $i$ are instructed to travel at speed $\mu_i$. 
Based on the SUMO simulation, the position of each vehicle
is then updated within ns-3.

The mobility information is relayed between ns-3 and SUMO through the Python
engine and the TraCI Python library. 
The ns-3 simulator and the Python
engine interact  through the \path{zmq} message-passing framework, using the client libraries available for both Python and C++. The communication between SUMO  and
the Python engine, instead, takes place through the TraCI protocol.  

In SUMO, flows include a mixture of different vehicle types, namely cars (SUMO class \path{passenger}, 85\% of all vehicles), trucks (class \path{truck}, 10\% of vehicles), and buses (class \path{coach}, 5\% of vehicles). Their target speed, acceleration, and driving aggressiveness values are left to the SUMO default, subject to a global speed limit of 50~km/h, as it is common in urban areas. The simulation step size of SUMO, which also determines the frequency of position updates in ns-3, is set to 10~ms.

In ns-3, LTE (provided by the LENA module) is used for the
communication between vehicles and infrastructure, while WAVE
(in the default ns-3 distribution) is used for
V2V communication, including that required for
lane change. CAMs and DENMs are encoded as foreseen by the ETSI 302.637
standard for ITS. 
Upon receiving a DENM, vehicles take action immediately, which
corresponds to the case of autonomous vehicles. Human reaction times,
usually quantified in 1\,s~\cite{noi-vtm}, could be easily accounted for.

\section{Numerical Results}
\label{sec:results}

We now demonstrate how our system model and the BH algorithm can be
exploited, with reference to our two applications of
assisted driving, namely, lane-change and navigation. Such
applications have different time and geographical scales, thus,
the fact that BH is effective in both cases demonstrates its
flexibility and generality.

Below, we first use a small-scale, yet representative, scenario, in
order to easily visualize the parameters and variables involved in our
decisions and better understand their mutual influence (\Sec{small}). In such a
scenario, we show the excellent performance of BH and highlight the impact
of the path probabilities $p_w^k$ on the system performance. Then we move
to larger, more complex scenarios (\Sec{medium}--\Sec{large}). In
these 
cases, we use BH to obtain the global policies and, 
as described in \Sec{validation}, evaluate their effect on the  travel times in real-world, practical
scenarios. 
Finally, we present some results on the convergence of BH  (\Sec{convergence}).

{\bf Benchmark solutions.}
We benchmark our BH scheme against two alternative solutions. 
The first is the fully-distributed solution  
where vehicles only communicate with neighboring vehicles 
(as per~\cite{autonet30-commag}), exploiting only local information to make lane-change decisions.
More specifically, vehicles
change lane if there are fewer neighbors (detected through CAMs) than in the current one. 
The second is a centralized solution based on the matching between flows and paths~\cite{bertossi1987some}. 
Such a solution runs the matching algorithm in~\cite{chen2016conflict} on a bipartite graph where:
(i) vertices represent the paths and flows;
(ii) edges connect each flow with the paths it can take;
    (iii) {\em conflict edges}~\cite{chen2016conflict} join the paths sharing one or more queues.
The algorithm in~\cite{chen2016conflict} yields a near-optimal solution, 
matching each flow with the paths resulting in the shortest travel time, and accounting for the fact that paths may share one or more queues,
leading to higher congestion.

\subsection{Lane-change, small-scale scenario
\label{sec:small}}

We begin by considering a lane-change service in the small-scale
scenario described in \Fig{archi} and modeled in \Fig{scenario}. Therein, there are two incoming
flows, namely, flows~1 and~2: the former is associated with one path
only, the latter with two paths. These two paths are called \path{early}
and \path{late}, referring to the fact that vehicles turn north
(respectively) after the first road segment (i.e., $q_2$), or after the
second one (i.e., $q_4$). We set the normalized incoming rate to
$\Lambda^k=1$~for both flows (rates are
  normalized to the arrival rate of flow 1), and the
maximum normalized service rate to $\mu^{\max}_i=3$ for all road segments, except for $q_4$ that, owing to
the fact that vehicles therein need to slow down, has a 
  maximum  normalized service rate $\mu_4^{\max}=1.5$. Also, we set
  $\omega^k$ to 5 time units for both flows. 
In such a  scenario, the best values of $\mu_i$ coincide with
$\mu^{\max}_i$ for  all road segments, thus 
the decisions to make are summarized by the
variable~$p^2_{\text{late}}$ (indeed, flow~1 only has one path and
$p^2_{\text{early}}=1-p^2_{\text{late}}$). 

\begin{figure}
\centering
\includegraphics[width=.8\columnwidth]{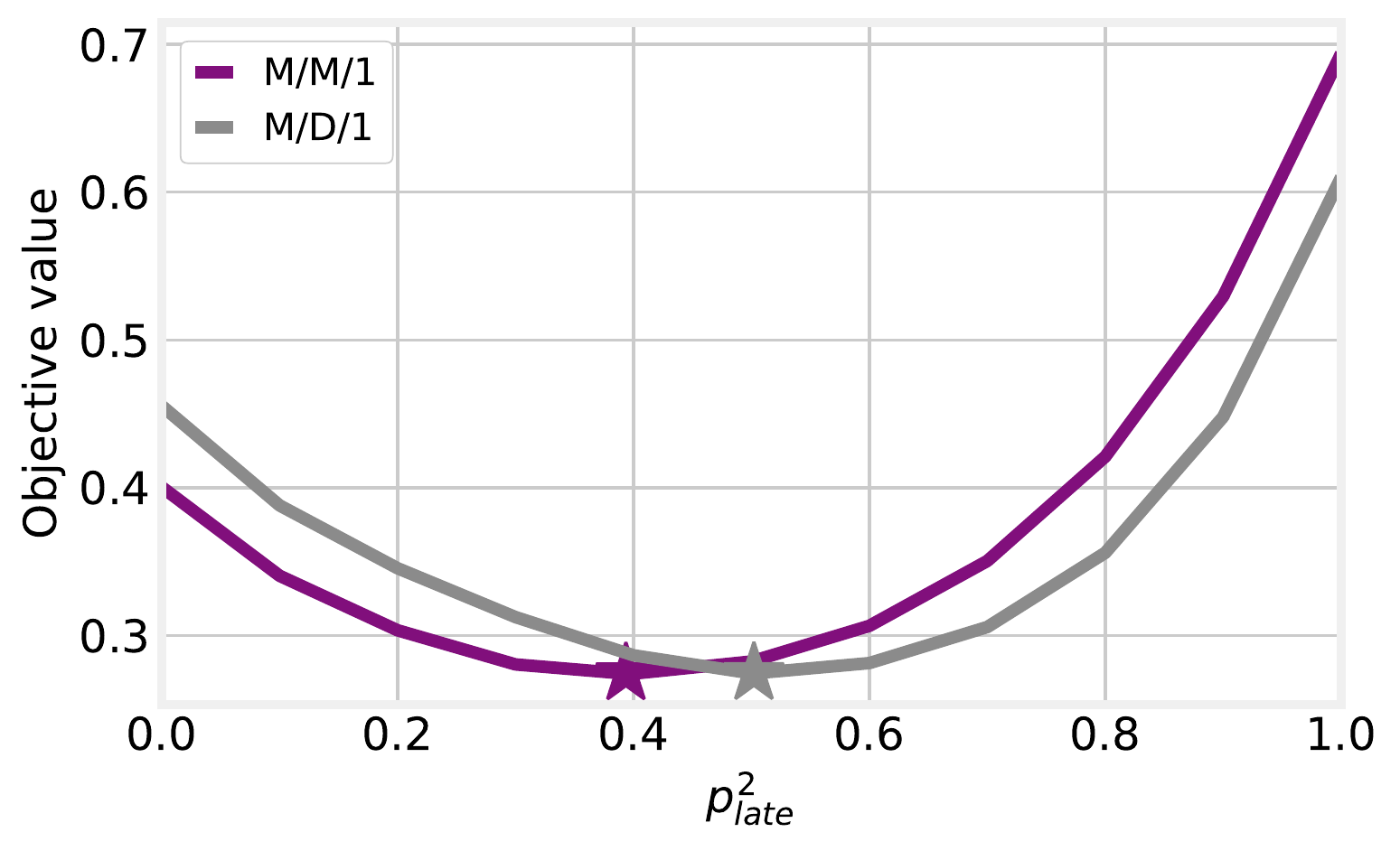}
\caption{
Small-scale scenario for lane change: value of the objective \Eq{obj-generic} as a 
function of~$p^2_{\text{late}}$, when road segments are modeled as $M/M/1$ (purple) 
or $M/D/1$ (gray) queues.
    \label{fig:small-comparison}
} %
\end{figure}

\begin{figure*}
\centering
\includegraphics[width=.32\textwidth]{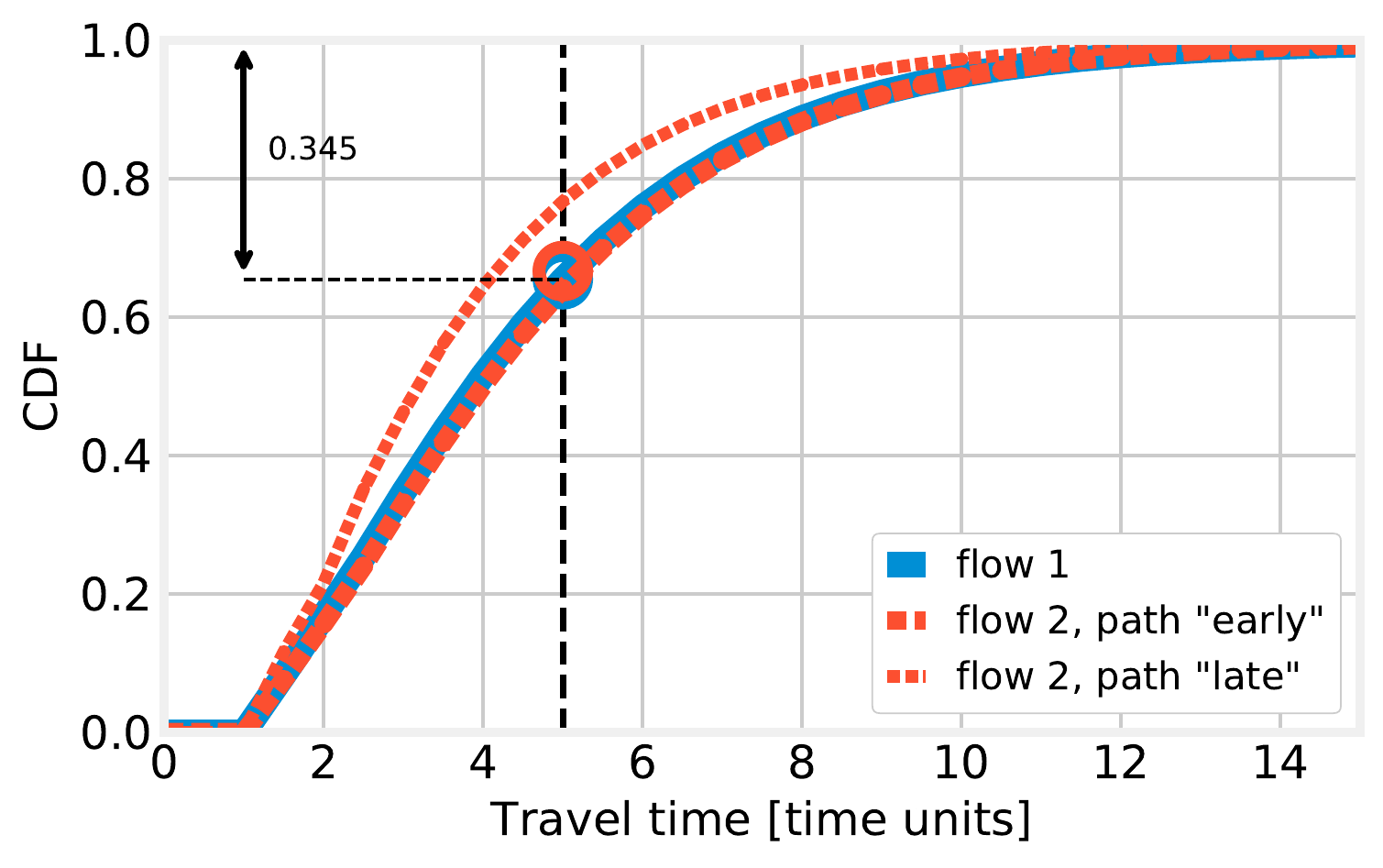}
\includegraphics[width=.32\textwidth]{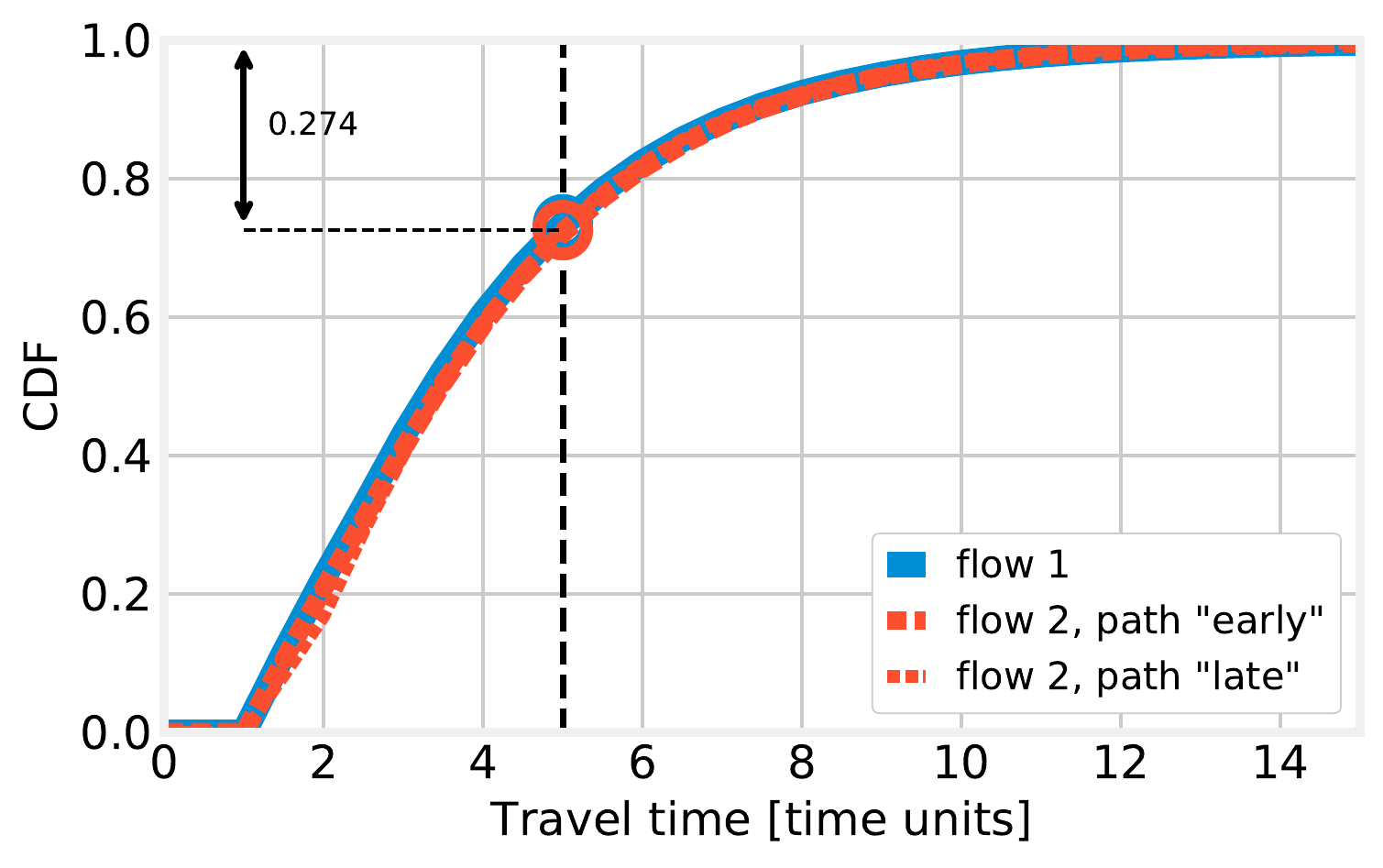}
\includegraphics[width=.32\textwidth]{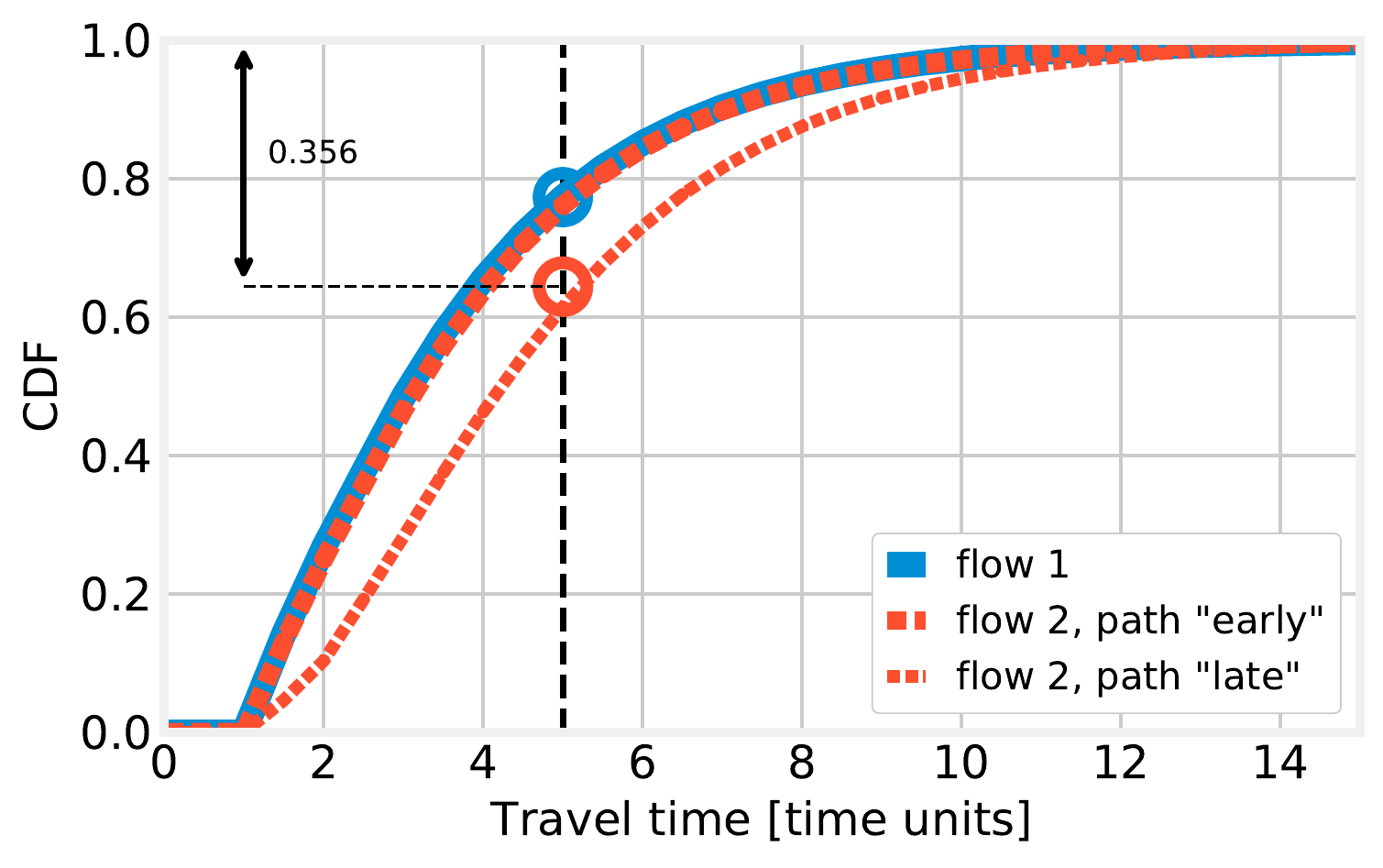}
\caption{
Small-scale scenario, lane-change,  $M/D/1$ queues: 
CDF of the per-path travel time when~$p^2_{\text{late}}=0.1$ (left), 
$p^2_{\text{late}}$~takes 
 its optimal value, as identified by both brute force and BH (center), and $p^2_{\text{late}}=0.9$ (right).
    \label{fig:small-cdfs-md1}
} %
\end{figure*}

\begin{figure}
\centering
\includegraphics[width=0.75\columnwidth]{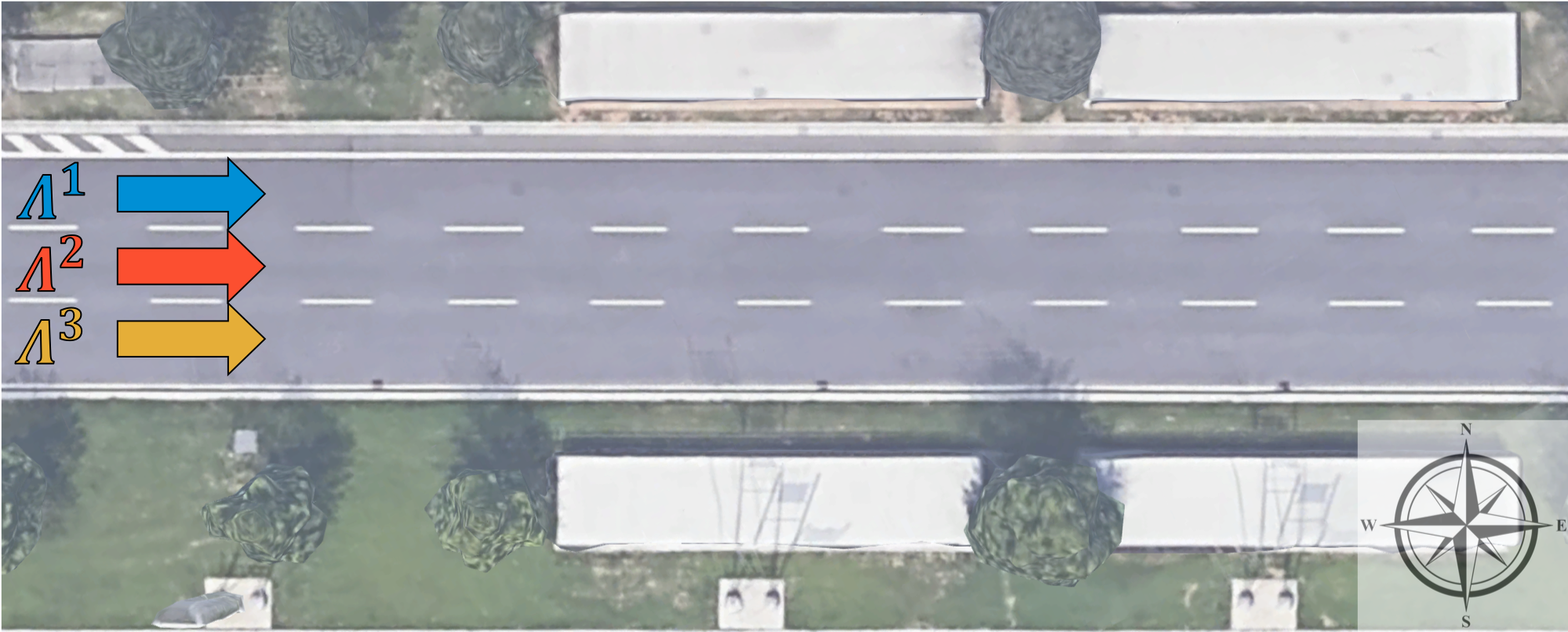}\\
\includegraphics[width=1\columnwidth]{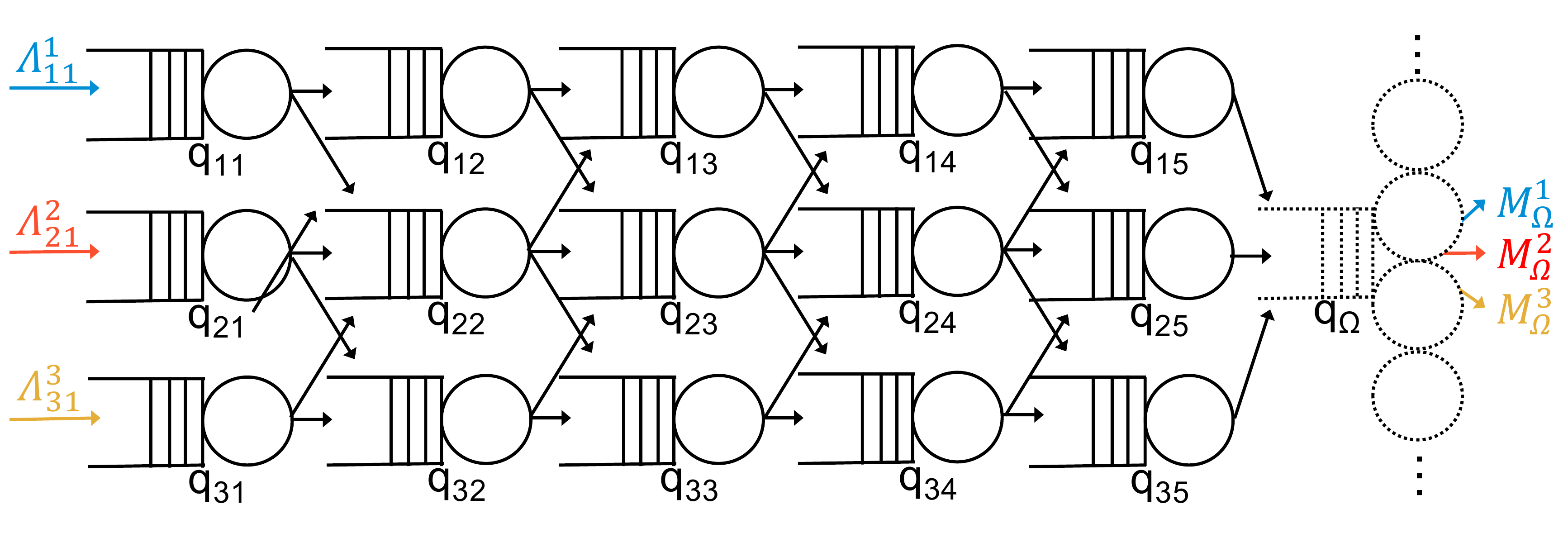}
\caption{
Medium-scale scenario for lane change: real-world
topology (top) and queue representation (bottom).
    \label{fig:scenario-medium}
} %
\end{figure}

\begin{figure*}
\centering
\includegraphics[width=.32\textwidth]{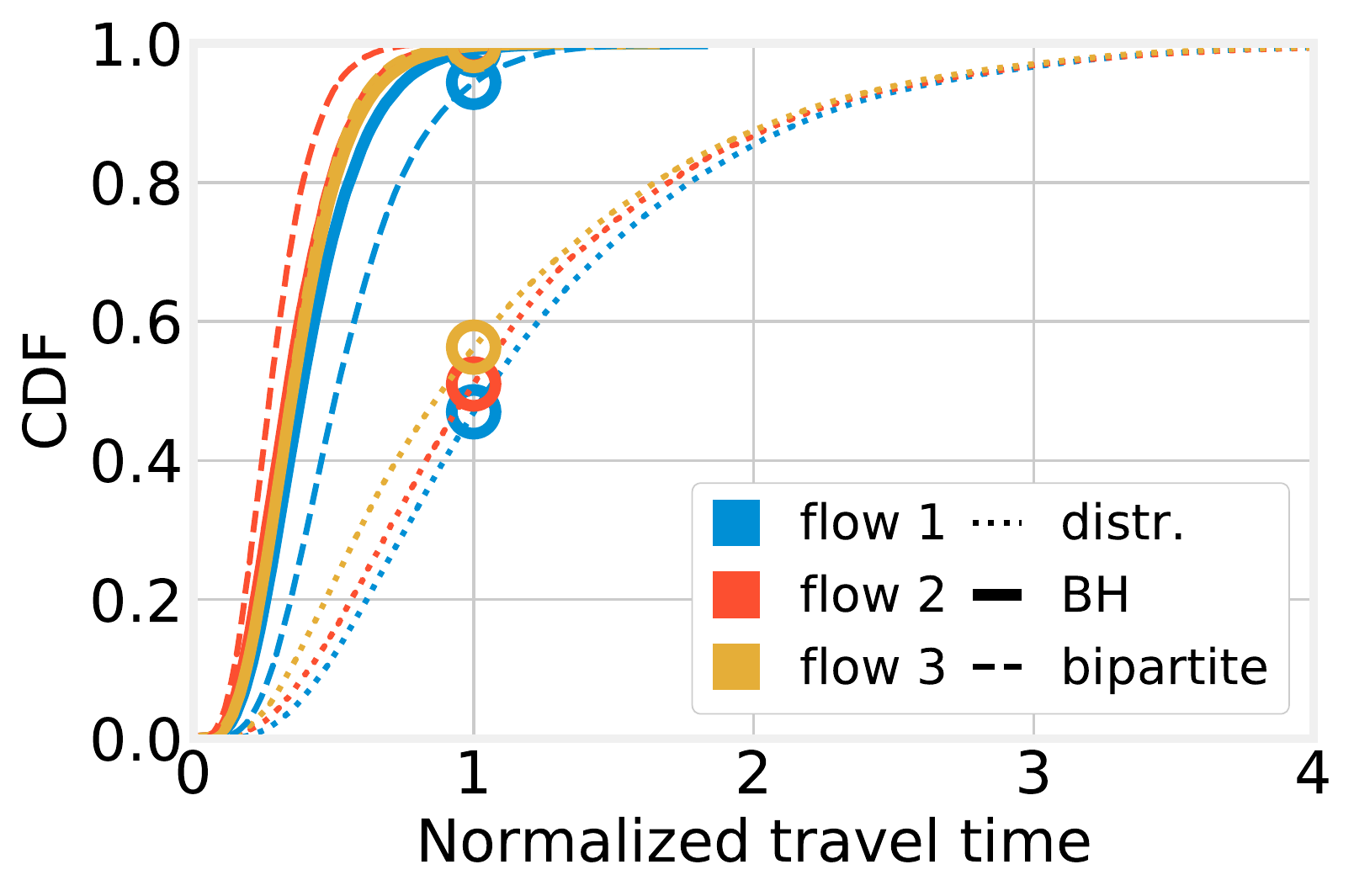}
\includegraphics[width=.32\textwidth]{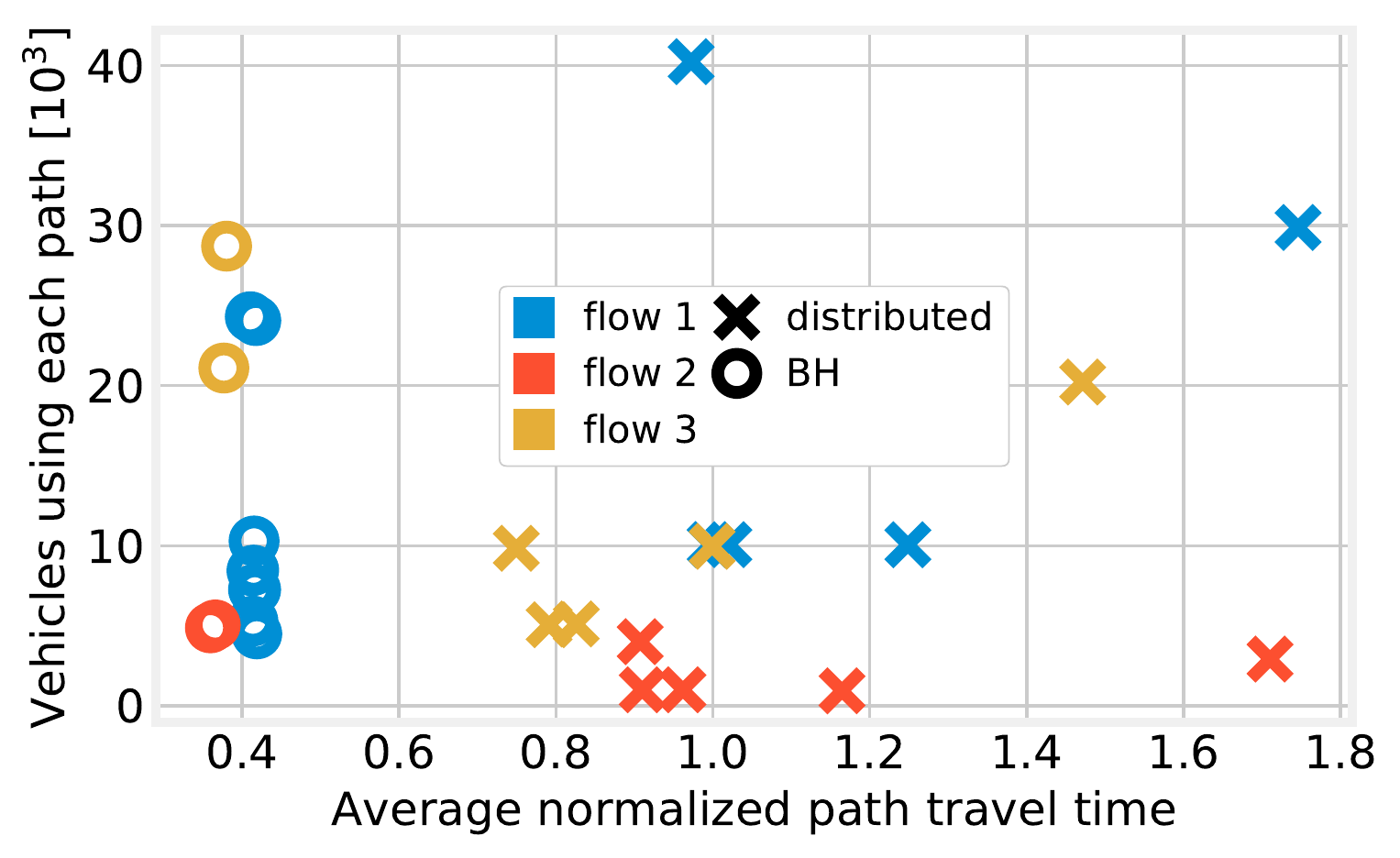}
\includegraphics[width=.32\textwidth]{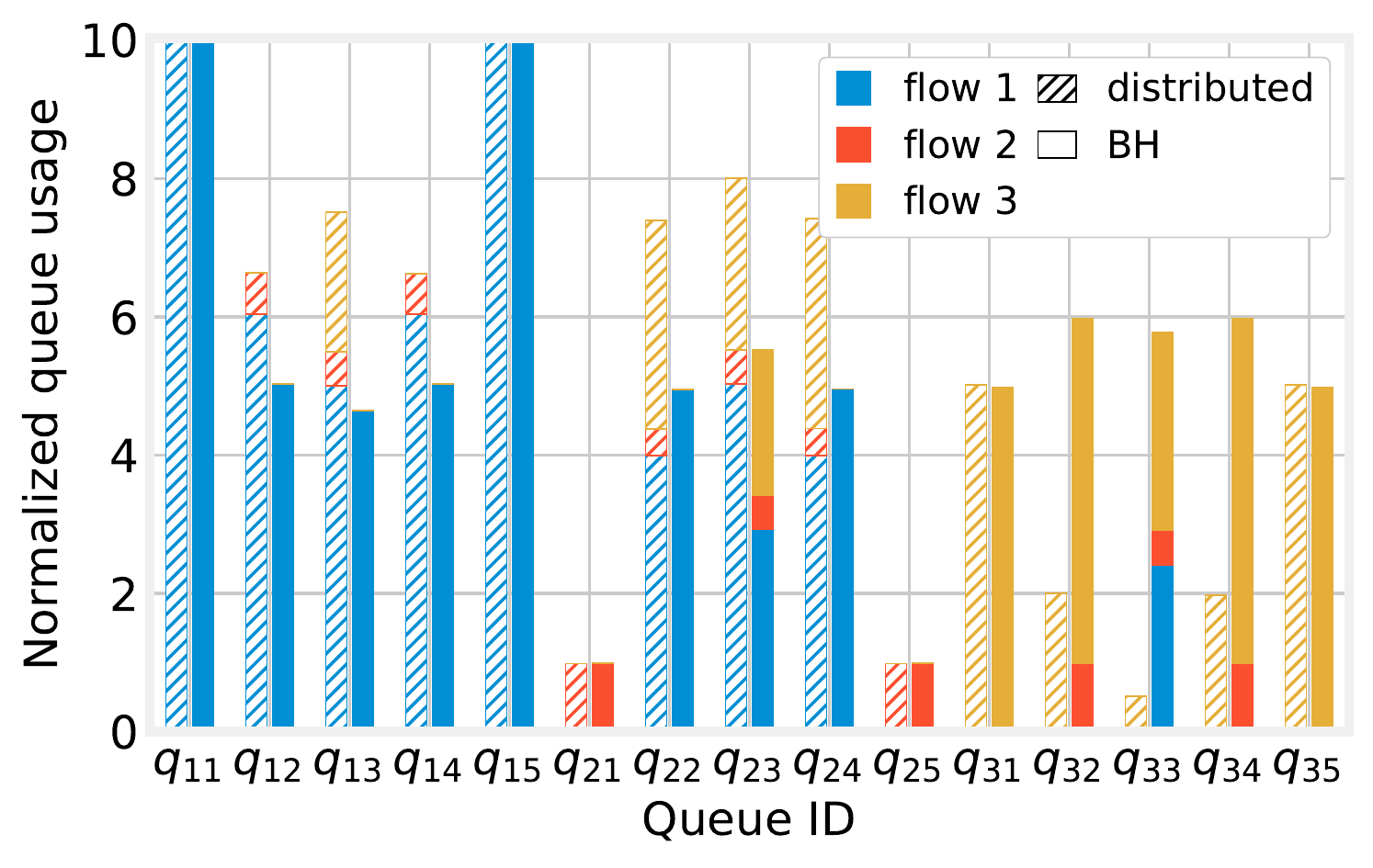}
\caption{
Medium-scale scenario, lane-change: CDF of per-flow travel
times (left); relationship between per-path travel times and number of
vehicles using it (center); number of vehicle per flow using each
queue (right).
    \label{fig:medium}
} %
\end{figure*}

\begin{figure}
\centering
\includegraphics[width=0.75\columnwidth]{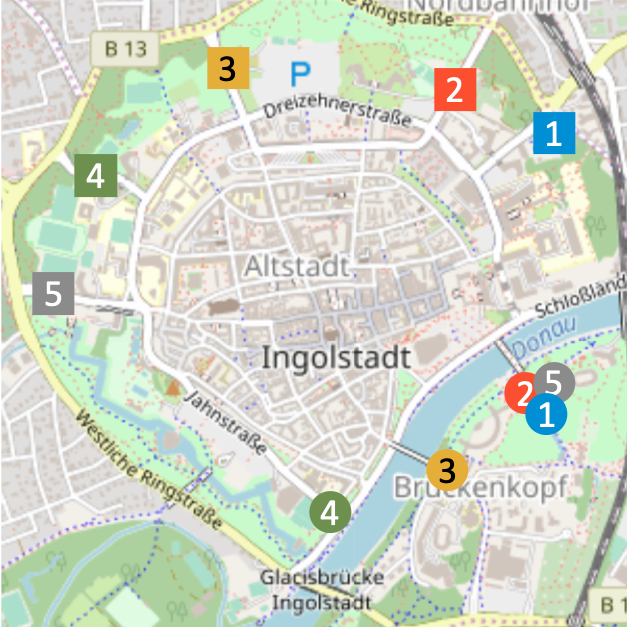}
\caption{
Large-scale scenario for navigation: the real-world topology we consider, with the origin (squares) and destinations (circles) of the flows.
    \label{fig:large-scenario}
} %
\end{figure}

The first, high-level question we seek to answer concerns the
relationship between the variable~$p^2_{\text{late}}$ and the value of
the objective function \Eq{obj-generic}. \Fig{small-comparison} shows
that it is advantageous to split the vehicles of flow~2 more or less
evenly between its two possible paths.

The third message conveyed by \Fig{small-comparison} is about
the flexibility of our approach:  the purple
curve is obtained by modeling road segments as $M/M/1$ queues, 
using the closed-form expressions in \Sec{sub-special}, while
the gray curve is obtained by using $M/D/1$ queues. Since there is
{\em no closed-form expression} of the sojourn time distribution in
$M/D/1$ queues, we 
implemented the approximate formula presented in~\cite{md1-times}, and
solved numerically integrals and convolutions. In spite of
the very significant differences with respect to the $M/M/1$ case, our
approach and BH  {\em work with no changes} in both
cases: our solution strategy does not depend on any specific service time distribution, 
and does not require such a distribution to have a closed-form expression.

Still for  the $M/D/1$ case, \Fig{small-cdfs-md1} shows the distribution of the per-path travel times, along with a graphical interpretation of the~$\omega^k$, $\delta_w(\omega^k)$, and~$\delta^k(\omega^k)$ quantities.
Each curve corresponds to a path, and its color represents the flow
each path belongs to (namely, blue: flow 1, red: flow 2). Firstly, we
observe  that, as the intuition would suggest, a higher value
of~$p^2_{\text{late}}$, i.e., sending more vehicles to path
\path{late} of flow~2, results in longer travel times for that path,
and shorter ones for path \path{early}. Secondly, the black vertical line in all
plots corresponds to  the value of the target travel time $\omega^k$;
ideally, one would like all CDFs to be at the left of such a line. 
 The intersections
between the CDFs and the vertical line represent the
probability that  vehicles belonging to a flow take a {\em path} whose travel time
does not exceed~$\omega^k$, i.e., $(1-\delta_w(\omega^k))$. 
Instead, the circles noted on the plots represent the {\em flow}-wise performance~$\delta^k(\omega^k)$:
as specified in \Eq{prob-k}, such a
quantity is defined as a weighted sum of the $\delta_w(\omega^k)$
values, thus, $\delta^2(\omega^2)$ is close
to~$\delta_\text{early}(\omega^2)$ when most vehicles take path
\path{early} (left plot), and close
to~$\delta_\text{late}(\omega^2)$ in the opposite case (right plot). 

The lowest circle, and the numerical values reported in the plots, correspond to the largest value
of~$\delta^k(\omega^k)$, i.e., the value of the objective function in \Eq{obj-generic}: 
intuitively, reducing the value of the objective function in \Eq{obj-generic}
corresponds to pushing up  such a circle as high as possible. 
Note that the flow with the highest~$\delta^k(\omega^k)$ changes for different values of~$p^2_\text{late}$: 
it is flow~1 in the left plot, and  flow~2 in the right one.
In the center plot, corresponding to the best solution as identified by both brute force and BH, 
all paths exhibit roughly the same travel time distribution. This is consistent with the intuition we gather from Thm.\,\ref{thm:allequal} and leverage on the design of the BH algorithm: similar path travel times result in better performance. Importantly, the hypothesis of Thm.\,\ref{thm:allequal} does not hold in this scenario -- and, indeed, path travel times are not {\em exactly} the same -- nonetheless, the intuition we gather from it is sill valid.

\subsection{Lane-change, medium-scale scenario}
\label{sec:medium}

We now move to a medium-scale, lane-change scenario, with three
vehicular flows, numbered from 1 to~3, traveling across a three-lane
road, as depicted in \Fig{scenario-medium}. 
Traffic flows originate at the beginning of one of the three lanes 
and can terminate at the end of {\em any} lane, not necessarily the
one they started at. Since, in our model, flows must have a unique
destination, we introduce a further, fictitious queue~$q_\Omega$, and make all
flows terminate there, as depicted in
\Fig{scenario-medium}(bottom). While all other queues are modelled as
$M/M/1$, $q_\Omega$~is an  
  $M/D/\infty$ queue
where the sojourn time coincides with the deterministic service
time and can  easily be discounted from the per-path
and per-flow travel times.

In this scenario, we consider the lane-change service
and translate the BH policies into instructions to vehicles, following the
methodology described in \Sec{validation}. Note that the
challenge here  is not due to an obstacle on the road (which is not
present any longer), but to the
significant difference among the incoming flows, namely, $\Lambda^1=10$,
$\Lambda^2=1$, $\Lambda^3=5$. The maximum service rate 
of all queues is $\mu^{\max}=15$, while 
the target travel time is  set to $\omega^k=1$ for all
flows.  Note that rate values are normalized
to $\Lambda^2$, while time values are normalized to $1/\Lambda^2$.

\Fig{medium}(left) presents the CDF of the vehicle travel time for
each flow when decisions are made by BH (solid lines), or by the
distributed 
decision scheme~\cite{autonet30-commag} (dotted lines), 
or the bipartite-matching strategy.
 BH yields
much shorter travel times; furthermore, the difference between
the flows, which is evident when decisions are made in a distributed manner, almost
disappears.
The bipartite-matching strategy yields better performance than the distributed one -- which, intuitively, is due to the fact that such a strategy uses global information. However, it is unable to match BH due to its inability to break flows across multiple paths, like BH does.

An explanation of the difference between BH and the distributed solution is presented in
\Fig{medium}(center). Each marker in the plot corresponds to one of
the $|\Wc|=35$~existing paths, and its position along the x- and
y-axes is given, respectively, by the average travel time over the
path and the number of vehicles using it; crosses correspond to distributed
decisions, circles to BH. We can observe that, when decisions are made
in a distributed fashion, the usage and path travel time change considerably, and the
corresponding markers are spread throughout the plot. On the contrary,
with BH, the paths have travel times that are not only shorter, but also very similar to each other -- hence, the corresponding markers are clustered together, and often overlapping. This is a consequence of how BH chooses the flows and paths to act upon (\Line{kstar} and \Line{wstar} in \Alg{bh}), which in turn reflects the min-max nature of objective \Eq{obj-generic}: traffic is always moved from more-crowded paths to less-crowded ones.

\Fig{medium}(right) summarizes how each queue is used by every
flow. By comparing hatched bars (distributed decisions) to solid bars
(BH), we observe how BH moves a more significant
fraction of flow~1's traffic away from the northernmost lane towards
the center (and, also, southern) ones. 
Counterintuitively enough, the 
traffic of flow~3 is scarcely moved to the center lane (see $q_{2x}$), which is
taken  by vehicles of flow~1; rather, a significant portion of
flow~2 is moved to the southern lane to make room for flow 1 in the
center lane. 
Such behaviors are unlikely to emerge as a combination of distributed decisions; instead, our comprehensive approach and the BH algorithm are able to identify and enforce them.

\Fig{medium}(right) also explains why, when distributed  decisions are made,
the travel times of the lower-rate traffic flows (2 and~3) are quite long and similar to those of flow~1 (see \Fig{medium}(left)). The reason is that such flows are routed through the center lane, which is also used by vehicles of flow~1, thereby increasing the congestion and travel times of all flows. It also explains how it is possible for BH to reduce the travel times for {\em all} flows, contrary to the intuition that removing congestion from one flow unavoidably means increasing it for some other. This is connected with the well-known notion that, in queuing systems, the total {\em traffic} is conserved but the sum of sojourn times is not, since sojourn times increase much faster than linearly (hyperbolically, in $M/M/1$ queues) with queue utilization.

\subsection{Navigation service, large-scale scenario \label{sec:large}}

\begin{figure*}[ht!]
\centering
\includegraphics[width=.32\textwidth]{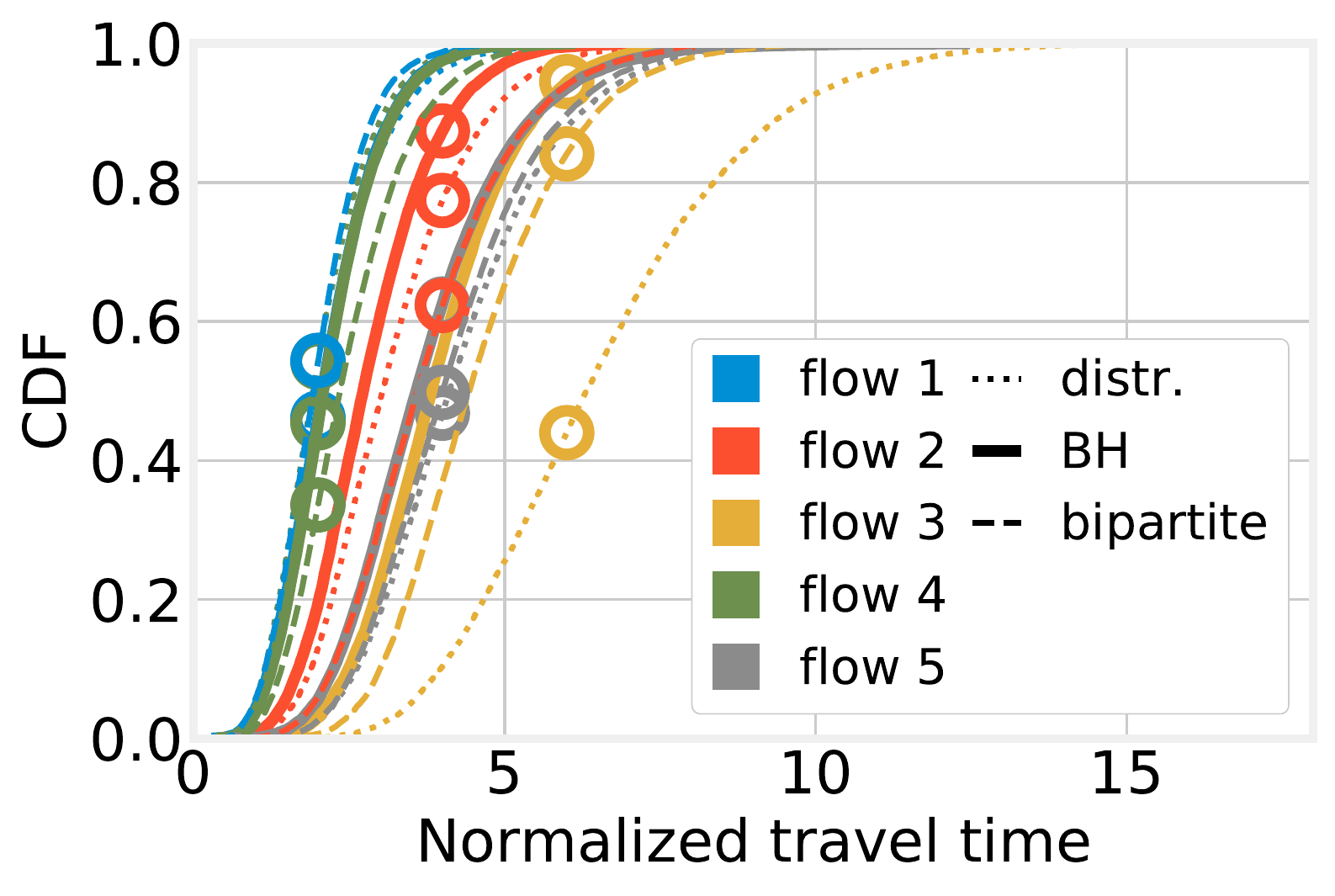}
\includegraphics[width=.32\textwidth]{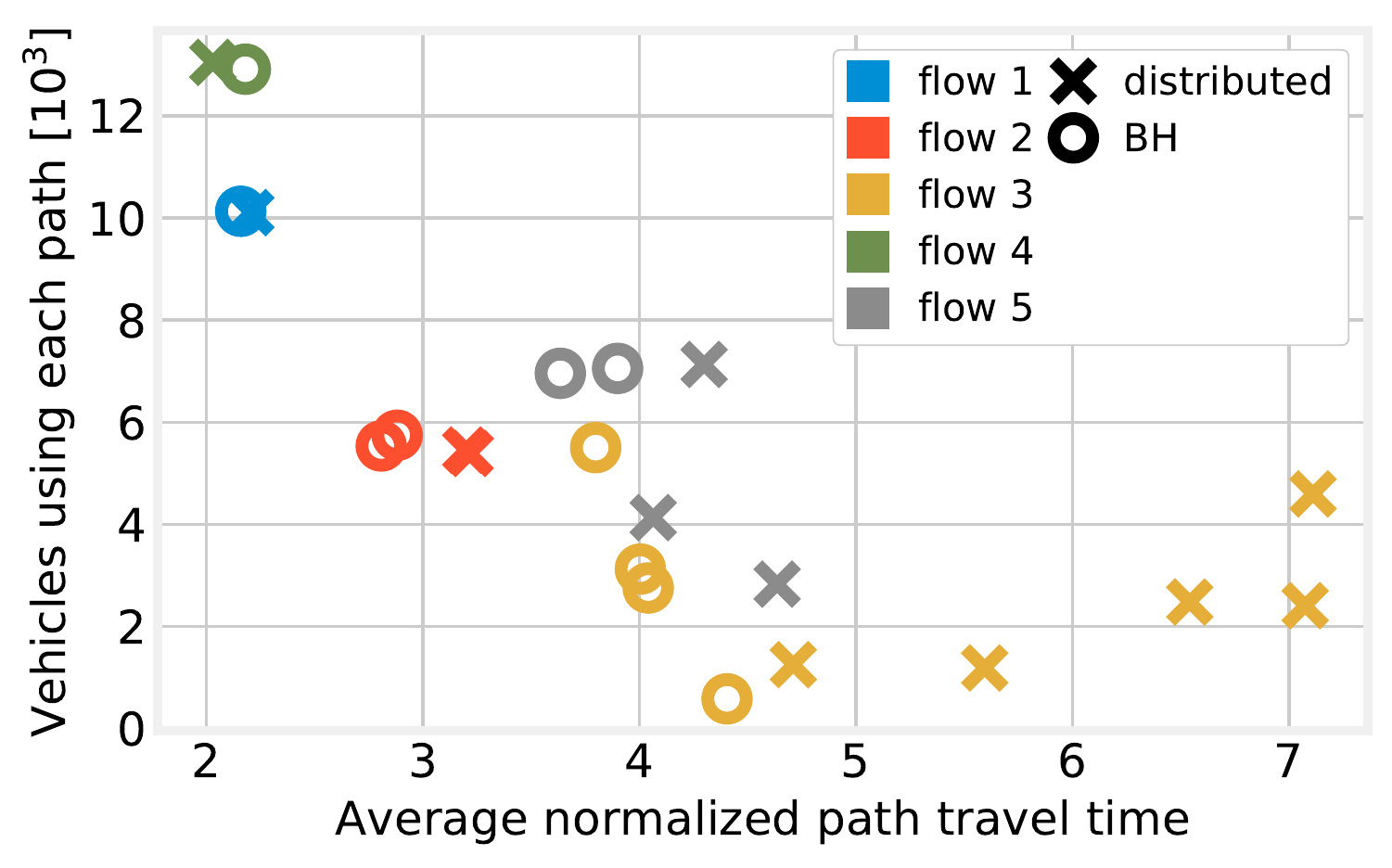}
\includegraphics[width=.32\textwidth]{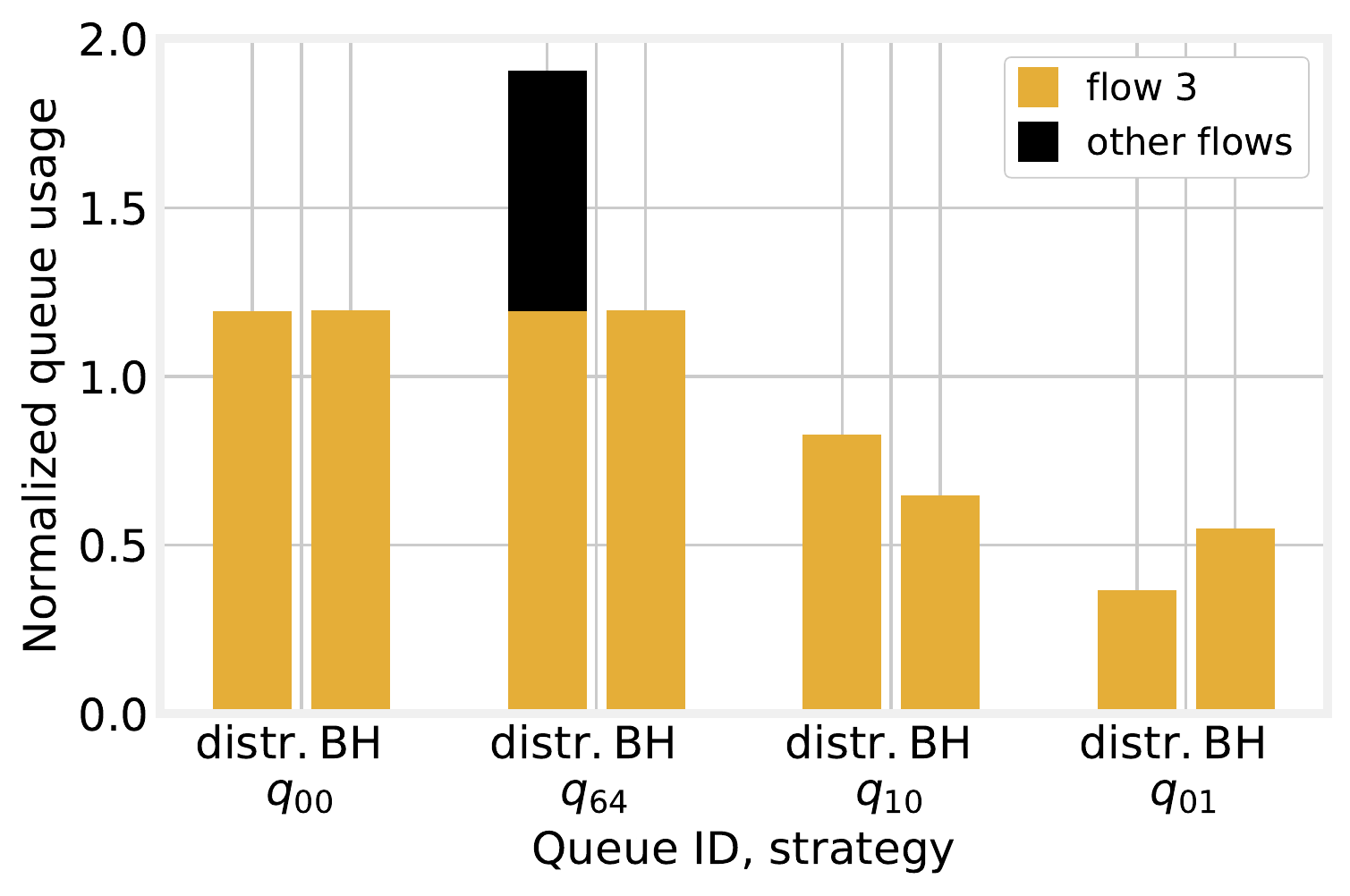}
\caption{
Large-scale scenario, navigation: CDF of per-flow
travel times (left); relationship between per-path travel times and
number of vehicles using it (center), traffic on the queues most used
by flow~$3$ (right). 
    \label{fig:large}
} %
\end{figure*}

Here we still evaluate the BH performance under practical conditions,
but we move to a large-scale scenario where a total of $|\Kc|=5$~flows
travel across a road topology including~$|\Qc|=49$ road segments
modelled as $M/M/1$ queues, 
yielding $|\Wc|=15$~paths. Both the
road topology and the intensity of traffic flows are extrapolated from
the real-world mobility trace in~\cite{trace} (see \Fig{large-scenario}),
obtaining a many-to-many-to-many correspondence between flows, paths and queues. %

In such a scenario, we provide a navigation service, aiming at guiding
all vehicles from the origin of their trip to their
destination, within their target time $\omega^k$. In
particular, we set the following normalized $\omega^k$-values:
$\omega^1=2$, $\omega^2=4$, $\omega^3=6$, $\omega^4=2$, $\omega^5=4$.
The source and destination of flows, as well as their intensity (i.e., the number of vehicles belonging to each of them) come from the real-world trace~\cite{trace}. 
Notice how flows are fairly diverse in length and parts of the city to traverse: 
some, like flow~1 and flow~4, only concern suburban areas of the city; others, 
like flow~3, need to cross the city center. 
Similarly, some flows (e.g., flow~1 and flow~2) are more likely than others to 
use the same road segments.

\Fig{large}(left) summarizes the CDF of the travel times for each
flow: solid lines correspond to  decisions made with BH, while dashed
lines correspond to distributed decisions made by vehicles based on
shortest-path routing. 
Interestingly, decisions made by BH result in substantially shorter travel times for the busiest flows, namely, flow~3 (yellow lines), flow~5 (gray lines), and flow~2 (red lines). Flow~1 (blue lines) is virtually unchanged, while the least congested flow, flow~4 (green lines), experiences slightly longer travel times. As discussed earlier, BH reduces the travel times of congested flows, without increasing -- or, as in this case,  increasing to a very limited extent -- the others.
The matching-based solution yields trip time similar to BH (as both solutions employ global information) but slightly longer (as BH is able to accurately split flows across paths).

\Fig{large}(center) provides more details on how such an improvement
is achieved. The performance of flows~3 and 5 is improved mostly by
avoiding the most crowded paths (rightmost yellow markers), and
routing instead more vehicles on less-crowded ones (leftmost
markers). This comes at the cost of increasing the load on some of the
queues shared with flow~4 (green markers): the circle representing its
only path is indeed further right than the corresponding cross. The
markers of flow~1 overlap, which is consistent with the fact that, as shown in \Fig{large}(left), the flow performance does not change.

\Fig{large}(right) focuses on the busiest flow (flow~3), and the
queues serving the highest number of vehicles thereof. For each queue,
yellow bars represent the number of vehicles of flow~3 traversing it,
and black ones the number of vehicles of all other flows. From the
figure, it is possible to observe two of the main ways in which BH
improves flow~3's performance. The first is clearly exemplified by
queue~$q_{64}$: if a queue is heavily used by flow~3, then vehicles of
other flows are directed elsewhere, thereby lowering the travel 
times.
In other words, BH improves the travel time of the less fortunate flows (as per \Eq{obj-generic}), by preventing multiple flows from crowding the same road segments.

The second can be observed in queues~$q_{10}$ and~$q_{01}$: 
as far as possible, BH ensures that all queues have the same load. 
This is again consistent with the insight provided by  Thm.\,\ref{thm:allequal}, 
that similar loads are associated with shorter travel times.

{\bf Summary.}
As reported in \Tab{summary1} and \Tab{summary2}, the BH algorithm can consistently outperform its alternatives, 
in terms of vehicle trip times and queue occupancy. As shown in \Tab{summary1}, the average trip time drops 
by 66\% in the lane-changing application (medium scenario), and by 20\% in the navigation one (large scenario).

This highlights the validity of the BH approach, which can split flow across paths (unlike the bipartite-matching one), 
while at the same time accounting for global information and the potentially far-reaching consequences of local decisions 
(unlike the distributed approach).

\subsection{Convergence speed \label{sec:convergence}}

Last, we focus on the convergence of BH and compare it against the
iterative algorithm BFGS~\cite{bfgs}, as implemented
in the \path{scipy} library. \Tab{convergence} reports the convergence
speed (expressed in number of objective function evaluations) of the
two algorithms.  BH consistently requires
fewer iterations than BFGS to converge, which confirms our
intuition that leveraging problem-specific knowledge and insights
results in faster convergence. It is also interesting to notice how
convergence in the medium-scale scenario 
 takes longer than in the large-scale one, owing to the
  higher number of paths (hence, of decisions to make) therein. 

\begin{table}[h]
\caption{
    Average trip time (across all vehicles) in the medium and large scenarios, in time units
    \label{tab:summary1}
}
\centering
\begin{tabular}{|l||r|r|r|}
\hline
Scenario & BH & bipartite & distributed\\
\hline\hline
Medium & 4.01 & 4.71 & 11.9\\
\hline
Large & 30.1 & 34.3 & 37.6\\
\hline
\end{tabular}
\end{table}

\begin{table}[h]
\caption{
    Average number of vehicles at each queue in the medium and large scenarios
    \label{tab:summary2}
}
\centering
\begin{tabular}{|l||r|r|r|}
\hline
Scenario & BH & bipartite & distributed\\
\hline\hline
Medium & 0.42 & 0.51 & 1.27\\
\hline
Large & 0.37 & 0.42 & 0.45\\
\hline
\end{tabular}
\end{table}

\begin{table}
\caption{
    Convergence speed  of  BH and BFGS
    \label{tab:convergence}
}
\centering
\begin{tabular}{|l|l||r|r|}
\hline
\multicolumn{1}{|c|}{Metric} & \multicolumn{1}{c||}{Scenario} & \multicolumn{1}{c|}{BH} & \multicolumn{1}{c|}{BFGS} \\ \hline\hline
\multirow{3}{*}{\begin{tabular}[c]{@{}l@{}}Convergence speed\\ {[}no. of objective\\ evaluations{]}\end{tabular}} & Small & 210 & 613 \\ \cline{2-4} 
 & Medium & 843 & 1039 \\ \cline{2-4} 
 & Large & 512 & 797 \\ \hline
\end{tabular}
\end{table}

\section{Conclusions}
\label{sec:conclusion}

We addressed the problem of providing connected vehicles and their drivers with effective
assistance services. We proposed an edge-based system including two logical entities, namely, 
policy maker and actualizer.
The former leverages global information, in order to make effective vehicle flow policies,  
while the latter translates such policies into instructions that the single vehicles should 
enact in the short time span. 
We modeled the system under study by leveraging  a queue-based
representation of the road topology and vehicles' behavior, which
allows for formulating an optimization problem that minimizes the vehicles' travel time. 
Then, motivated by the problem complexity and the need to address
large-scale scenarios, we presented 
a swift iterative algorithm providing optimal policies in linear time.
We assessed the performance of the proposed approach through a
full-fledged, realistic simulation
framework, showing the benefits of our solution in terms of 
vehicles'  travel times with respect to traditional distributed approaches.

\bibliographystyle{IEEEtran}
\bibliography{refs_short}

\end{document}